\documentclass[useAmain sequence,usenatbib]{mn2e}
\usepackage{graphicx,subfigure}
\usepackage{bm}
\usepackage{aalongtable}
\usepackage{rotating}
\def\aj{AJ}
\def\araa{ARA\&A}
\def\apj{ApJ}
\def\apjl{ApJL}
\def\apss{Ap\&SS}
\def\aap{A\&A}
\def\mnras{MNRAS}
\def\pasp{PASP}
\def\nat{Nature}
\def\iaucirc{IAU~Circ.}

\DeclareMathAlphabet{\mathsc}{OT1}{cmr}{m}{sc}
\def\testbx{bx}%
\DeclareRobustCommand{\ion}[2]{%
\relax\ifmmode
\ifx\testbx\f@series
{\mathbf{#1\,\mathsc{#2}}}\else
{\mathrm{#1\,\mathsc{#2}}}\fi
\else\textup{#1\,{\mdseries\textsc{#2}}}%
\fi}
\newcommand{\ha} {\mbox{H$\alpha$}}
\newcommand{\hb} {\mbox{H$\beta$}}
\newcommand{\hg} {\mbox{H$\gamma$}}

\newcommand{\Feiia} {[\ion{Fe}{ii}]}
\newcommand{\Feii} {\ion{Fe}{ii}}

\newcommand{\Caiia} {[\ion{Ca}{ii}]}
\newcommand{\Coiia} {[\ion{Co}{ii}]}

\newcommand{\Caii} {\ion{Ca}{ii}}

\newcommand{\Baii} {\ion{Ba}{ii}}
\newcommand{\Nai} {\ion{Na}{i}}

\newcommand{\Mgia} {\ion{Mg}{i}]}
\newcommand{\Mgii} {\ion{Mg}{ii}}

\newcommand{\Hii} {\ion{H}{ii}}

\newcommand{\Tiii} {\ion{Ti}{ii}}
\newcommand{\Hei} {\ion{He}{i}}

\newcommand{\Nii} {\ion{N}{ii}}
\newcommand{\Oia} {[\ion{O}{i}]}
\newcommand{\Oiia} {[\ion{O}{ii}]}
\newcommand{\Oi} {\ion{O}{i}}

\newcommand{\Ariii} {[\ion{Ar}{iii}]}

\newcommand{\SiII} {\ion{Si}{ii}}

\def\sn{SN 2007uy}

\newcommand{\eg}{{\textrm e.g.}}
\newcommand{\ie}{{\textrm i.e.}}

\newcommand{\ebv}{\mbox{$E(B-V)$}}
\newcommand{\ew}{\mbox{$EW$}}

\newcommand{\degree}{\mbox{$^\circ$}}
\newcommand{\msun}{\mbox{M$_{\odot}$}}

\newcommand{\kms}{\mbox{$\rm{\,km\,s^{-1}}$}}

\newcommand{\nickel}{\mbox{$^{56}$Ni}}

\newcommand{\mum}{\mbox{$\mu{\rm m}$}}

\begin{document}
\title[SN 2007uy in NGC 2770]
{SN 2007uy $-$ metamorphosis of an aspheric Type Ib explosion}
\author[Roy et al.]
{Rupak Roy$^1$ \thanks{e-mail: roy@aries.res.in, rupakroy1980@gmail.com},
 Brijesh Kumar$^{1}$, Justyn R. Maund$^2$\thanks{Royal Society Research Fellow},
 Patricia Schady$^3$,
\newauthor
 Felipe Olivares E.$^3$, Daniele Malesani$^4$, Giorgos Leloudas$^{4,5}$, Sumana
 Nandi$^1$,
\newauthor
 Nial Tanvir$^6$, Dan Milisavljevic$^7$, Jens Hjorth$^4$, Kuntal Misra$^1$,
 Brajesh Kumar$^1$,
\newauthor
 S. B. Pandey$^1$, Ram Sagar$^1$, H. C. Chandola$^8$\\
\\
$^{1}${Aryabhatta Research Institute of Observational Sciences, Manora Peak, Nainital, 263 002, India}\\
$^{2}${Astrophysics Research Centre, School of Mathematics and Physics, Queen’s University Belfast, Belfast BT7 1NN, UK}\\
$^{3}${Max-Planck-Institute f\"{u}r extraterrestrische Physik, Giessenbachstra$\ss$e 1, 85748 Garching, Germany}\\
$^{4}${Dark Cosmology Centre, Niels Bohr Institute, University of Copenhagen, Juliane Maries vej 30, 2100 Copenhagen $\O$, Denmark}\\
$^{5}${The Oskar Klein Centre, Department of Physics, Stockholm University, Albanova University Centre, 10691, Stockholm, Sweden.}\\
$^{6}${Department of Physics and Astronomy, University of Leicester, Leicester, LE1 7RH, United Kingdom}\\
$^{7}${Department of Physics and Astronomy, Dartmouth College, Hanover, NH 03755}\\
$^{8}${Department of Physics, Kumaun University, Nainital, India}\\
} 

\date{Accepted ??; Received ??}

\pagerange{\pageref{firstpage}--\pageref{lastpage}} \pubyear{}

\maketitle

\label{firstpage}


\begin{abstract}
 {The supernovae (SNe) of Type Ibc are rare and the detailed
 characteristics of these explosions have been studied only for a few events.
 Unlike  Type II SNe, the progenitors of Type Ibc have never been detected in
 pre-explosion images. So, to understand the nature of their progenitors and the
 characteristics of the explosions, investigation of proximate events are
 necessary. Here we present the results of multi-wavelength observations of 
 Type Ib \sn\ in the nearby ($\sim$ 29.5 Mpc) galaxy NGC 2770. Analysis of the
 photometric observations revealed this explosion as an energetic event with
 peak absolute $R$ band magnitude $-18.5\pm0.16$, which is about one mag
 brighter
 than the mean value ($-17.6\pm0.6$) derived for well observed Type Ibc events.
 The SN is highly extinguished, \ebv\ = 0.63$\pm$0.15 mag, mainly due to
 foreground material present in the host galaxy.
 From optical light curve modeling we determine that about 0.3 \msun\
 radioactive $^{56}$Ni is produced and roughly 4.4 \msun\ material is ejected
 during this explosion with liberated energy $\sim 15\times10^{51}$ erg, 
 indicating the event to be an energetic one. Through optical spectroscopy, we
 have noticed a clear aspheric evolution of several line forming regions, but no
 dependency of asymmetry is seen on the distribution of $^{56}$Ni inside the
 ejecta.
 The SN shock interaction with the circumburst material (CSM) is clearly
 noticeable in radio follow-up, presenting a Synchrotron Self Absorption (SSA)
 dominated light curve with a contribution of Free Free Absorption (FFA) during
 the early phases. Assuming a Wolf-Rayet (WR) star, with wind velocity
 $\ga 10^3 {\rm km~s}^{-1}$, as a progenitor, we derive a lower limit to the
 mass loss rate inferred from the radio data as
 $\dot{M} \ga 2.4\times10^{-5}$ \msun\,~yr$^{-1}$, which is consistent with the
 results obtained for other Type Ibc SNe bright at radio frequencies.
 }
\end{abstract}


\section{Introduction}
\label{sec:int}
 Core-collapse supernovae (CCSNe) mark the end stage of the evolution of massive
 stars having initial masses greater than 8\msun\ \citep{2003ApJ...591..288H,
 RevModPhys.85.245}. Stars, more massive than 10\msun\ produce an Fe-core, and
 then expel their outer layers in a catastrophic explosion (e.g.,
 \citealt{2009ApJ...705L.138P}).
 Type Ib SNe are a class of CCSNe which are spectroscopically defined by the
 absence of hydrogen (H) and presence of helium (He) in their early phase
 optical spectra. It is generally accepted that Type Ib SNe along with Type Ic
 (absence of H and He) are  formed from evolved high mass progenitors
 like Wolf-Rayet (WR) stars which have liberated their outer shells through
 pre-SN winds. The outer shells of H \& He may also be stripped through mass
 transfer due to Roche-lobe overflow in a binary system. Type Ibc SNe are
 commonly referred to as ``Stripped-envelope Supernovae''
 \citep{1997ApJ...491..375C}. The progenitors of Type Ibc SNe have not been
 observed directly in pre-explosion high resolution images. There are only a few
 cases, like SNe 2000ds, 2000ew, 2001B and 2004gt where upper-limits on the flux
 of the Type Ibc progenitors have been reported \citep{2005MNRAS.360..288M,
 2005ApJ...630L..33M, 2013arXiv1301.1975E}. After analyzing
 the evolutionary models of massive He stars and comparing the results with
 Galactic WR stars, it has been found that during pre-SN stage, progenitors of
 Type Ibc SNe have surface properties that resemble those of hot Galactic WR
 stars of WO sub-type, which are visually faint, M$_V$ $\approx -1.5{~\rm
 to}-2.5$ mag, despite of a high bolometric luminosity, L/L$_\odot = 5.6-5.7$
 \citep{2012A&A...544L..11Y}. Detection of Type Ic progenitors is even
 more challenging than that of Type Ib. Monitoring of this class of SNe is
 crucial, not only to investigate the post-explosion phenomenon, but also to
 understand the pre-explosion properties of the progenitors.
 
 With a volumetric sample of nearby CCSNe discovered by the Lick Observatory
 Supernova
 Search (LOSS) program, \citet{2011MNRAS.412.1522S} found that only about 7\% of
 total nearby CCSNe are of Type Ib. Furthermore they argued that most of the
 Type Ib are formed through mass transfer in a binary system rather than through
 the progenitor's wind and that the true progenitors of SNe Ib must extend to a
 much lower range of initial masses than classical WR stars.

 Apart from the issues related to progenitors and the pre-explosion properties,
 there are many unresolved issues concerning the post-explosion properties. One
 of them is the explosion geometry. Spectropolarimetry has revealed CCSNe to be
 more aspheric in their inner layers with a preferred direction of explosion in
 comparison with thermonuclear events, which instead show asphericity in their
 outer shells
 \citep{2008ARA&A..46..433W}. Among CCSNe, Type IIP events show a moderate
 non-zero polarization in the plateau phase,
 whereas Type Ibc SNe are aspheric explosions with substantially higher
 intrinsic polarization from the beginning \citep{2001ApJ...550.1030W,
 2007MNRAS.381..201M}. The asphericity is explained as being due to clumpy
 structure or jet-like propagation of the ionizing radioactive Ni and Co source,
 concentrated at the inner portion of the spherically symmetric ejecta
 \citep{1992SvAL...18..168C, 2001AIPC..586..459H,2007MNRAS.381..201M}. In the
 light of mounting evidence of the asphericity of Type Ibc events, the double
 horned nature of the \Oia\ $\lambda\lambda6300,6364$ profile was explained as a
 jet/torus like structure \citep{2005Sci...308.1284M, 2008ApJ...687L...9M,
 2008Sci...319.1220M}. Though, alternatively, it can be explained as a combined
 effect of two \Oia\ distributions $-$ (i) a central, symmetric distribution of
 \Oia\,-rich ejecta, (ii) a clumpy or shell of \Oia\,-rich material traveling in
 the front facing hemisphere
 \citep[although this model also has some drawbacks; see][]{2010ApJ...709.1343M, 2009MNRAS.397..677T}. 
 Clearly, individual case studies
 are important to understand the preferred mechanism behind the ambiguous
 behavior of \Oia\ lines. 
 
\begin{figure}
\centering
\includegraphics[width=8.5cm]{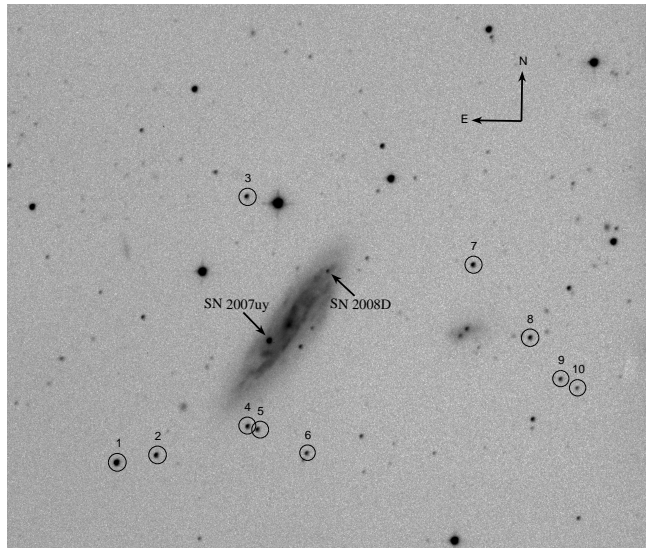}
\caption{Identification chart of SN 2007uy. The image is about 10 arcmin on a
 side taken in $V$-band with the 104-cm Sampurnanand telescope at ARIES,
 Nainital. The local secondard stars are numbered. North is up and east is to
 the left.}
\label{fig:snid}
\end{figure}

 SN 2007uy, a Type Ib event, was first discovered by Yoji Hirose on 2007,
 December 31.7 UT
 at an unfiltered magnitude 17.2 \citep{2008CBET.1191....1N} in the nearby
 galaxy NGC 2770 ($z \sim$ 0.007, distance $\sim 29.5$ Mpc). There are numerous
 intervening host galaxy \Hii\ regions along the direction of the SN
 \citep{2011AdSpR..47.1421G} and no object was found at the transient location
 in the red band DSS image observed between 1998 and 2000.
 \citet{2008CBET.1191....2B} reported the first optical spectroscopic
 observation of this event taken on 2008, January 3.40 UT, classifying this
 transient as a type Ib event similar to SN 2004gq.
 The discovery date corresponds to JD = 2454466.17. According to initial
 spectroscopy \citep{2008CBET.1191....2B} the explosion happened within 7 days
 prior to its discovery. Modeling of radio data (\S\ref{sec:modeling}) also
 indicates that the event was discovered within 4 days of the explosion. Here
 we adopt a conservative value of
 the explosion epoch to be 4 days prior the detection, \ie\, JD = 2454462.17.
 In rest of the work, if not mentioned, all the phases will be quoted with
 respect to this epoch. The field of \sn\ is presented in figure \ref{fig:snid}
 and the properties of the host galaxy and the SN are provided in Table
 \ref{tab:propgal}. A detailed
 description of the host can be found elsewhere \citep{2009ApJ...698.1307T}.
 SN 2007uy was also detected in radio bands at +10d
 \citep{2008ATel.1350....1S} with an X-band (8.46 GHz) flux density of
 0.29$\pm$0.03 mJy. The SN was also detected at X-ray wavelengths
 \citep{2008ATel.1368....1P}.

  \begin{table}
  \caption{Properties of the host galaxy NGC 2770 and SN 2007uy.}
  \label{tab:propgal} 
  \begin{tabular}{llc} \hline \hline
     \noalign{\smallskip}
      Parameters& Value& Ref.$^{a}$\\ 
     \noalign{\smallskip} \hline
    
     \noalign{\smallskip}
     \multicolumn{3}{l}{\bf NGC 2770:}\\
     Type& SABc& 1\\
     Position & $\alpha_{J2000} = 09^{\rm h} 09^{\rm m} 33\fs68$& 1\\
              & $\delta_{J2000} = +33\degr 07\arcmin 24\farcs7$&  \\
     Abs. magnitude& $M_{B}=-20.78$ mag& 1\\
     \\
     Distance& $D=29.5\pm1.8$ Mpc&  \S\ref{sec:DistExt}\\
     Scale& $1\arcsec \sim 136$ pc, ~~$1\arcmin \sim 8.2$ kpc&\\
     Distance modulus& $\mu \sim 32.4$ mag&\\
     \\
     Apparent radius&${r_{\mathrm{25}}}=1\farcm73\,(\sim 15.09\,\mathrm{kpc})$& 1\\
     Inclination angle&$\Theta_{\rm inc}= 82.3^\circ$&1\\
     Position angle& $\Theta_{\rm maj}=146.1^\circ$&1\\
     Heliocentric velocity& $cz_{\rm helio}=1947\pm2$\kms\,&1\\
     \\
     \multicolumn{3}{l}{\bf SN 2007uy:}\\
     Position & $\alpha_{J2000} = 09^{\rm h} 09^{\rm m} 35\fs40$& 2\\
              & $\delta_{J2000} = +33\degr 07\arcmin 09\farcs9$& \\
     \\
     Location& $20\farcs6$ E, $15\farcs5$ S&  2\\
     Deprojected radius&  $r_{\rm SN} = 42\farcs44$ ($\sim$ 5.3 kpc)& \S\ref{sec:DistExt}\\
     \\
     Discovery date (UT)& 31.7 December 2007&  2\\
                   & (JD 2454466.17)& \\ 
     Explosion epoch: & $\sim 4$ days prior to discovery& \S\ref{sec:int}\\ 
                   & (JD 2454462.17)& \\ 
     Total reddening      : & \ebv\ = $0.63\pm0.15$ mag& \S\ref{sec:DistExt}\\ 
   
     \noalign{\smallskip}
     \hline
  \end{tabular}
  \newline\newline
  $^{a}${(1) HyperLEDA - http://leda.univ-lyon1.fr; (2)
  \citet{2008CBET.1191....1N}.}
  \end{table}


 In this paper, we present photometric follow-up of \sn\ in near-ultraviolet
 (NUV), optical and near-infrared (NIR) bands and low resolution spectroscopic
 follow-up
 observations in optical. We summarize the observations and data reduction
 procedure in \S\ref{sec:obs}. We study the spectroscopic evolution in
 \S\ref{sec:specevo}. The distance and reddening have been estimated in
 section \S\ref{sec:DistExt}. In \S\ref{sec:lightcurve} photometric evolution is
 investigated, while the section \S\ref{sec:ColBol} presents the evolution of
 colour and bolometric light. The main physical parameters of the explosion and
 the characteristics of the progenitor are derived and discussed in
 \S\ref{sec:modeling} and \S\ref{sec:conclu}.

\section{Observations and data reduction}
\label{sec:obs}
\subsection{Near-ultraviolet, optical and near-infrared multiband photometry}
\label{sec:phot}
 The Ultraviolet Optical Telescope (UVOT) monitored the SN at 30 phases between
 +9d and +75d. The observations were carried out in $uvw2~(\lambda_c=2030$\AA$),
 uvm2~(\lambda_c=2231$\AA$), uvw1~(\lambda_c=2634$\AA$),
 u~(\lambda_c=3501$\AA$), b~(\lambda_c=4329$\AA$) {\rm~and}~v
 ~(\lambda_c=5402$\AA$)$ bands (see table \ref{tab:uvot_07uy} for details).
 We obtained the UVOT data
 from the $Swift$ Data Archive. The photometry was done using standard
 procedures using HEASOFT routines \footnote{The $Swift$ data have had bad
 pixels identified, mod-8 noise corrected, and have been transformed into FK5
 coordinates. We used the standard UVOT data analysis software distributed with
 HEASOFT 6.10 along with the standard calibration data. As long as the source
 had a count rate greater than 0.5cts\,s$^{-1}$, photometry was done using
 `uvotsource' with a standard circular aperture of radius 5$\arcsec$ and a
 circular background region with a radius of 15$\arcsec$. Below this threshold,
 a $3\farcs5$ radius was used and an aperture correction was applied. The
 background region was selected to have similar background properties to those
 at the location of the SN, and to be free of contaminating sources.} and
 it was calibrated to the UVOT photometric system following the procedure
 described in \citet{2008MNRAS.383..627P}. To remove the contribution of the
 underlying host galaxy, we measured the host galaxy flux at the position of the
 SN from late time UVOT observations, with no contribution from the SN. This
 additional flux was then subtracted from SN photometric measurements. For all
 observations the source was close to the centre of the field-of-view, and
 differences in the PSF (Point Spread Function) between observations were,
 therefore, negligible.

 The optical broadband Johnson $UBV$ and Cousins $RI$ follow-up observations
 were performed at 27 phases between $\sim$+16d to +130d using the 104-cm
 Sampurnanand Telescope (ST)+imaging camera\footnote{A 2048 $\times$ 2048 CCD
 camera having 24$\times$24 $\micron$ chip size and with a plate scale
 $0\farcs38$/pixel, mounted at the f/13 Cassegrain focus of the telescope was
 used for observation. The gain and readout noise of the CCD camera are 10
 electrons per analog-to-digital unit and 5.3 electrons respectively. To
 improve the signal-to-noise ratio, we performed the observations in a binned
 mode of 2$\times$2 pixel.} at Nainital, India and the 2.56 m Nordic Optical
 Telescope (NOT) + Andalucia Faint Object Spectrograph and Camera (ALFOSC) at La
 Palma, Spain. Several exposures, in the range
 100s to 300s, were taken and in order to increase the signal-to-noise-ratio
 (SNR), photometry was performed on the co-added frames. The pre-processing of
 raw data were performed through standard data reduction software
 {\it IRAF}\footnote{{\it IRAF} stands for Image Reduction and Analysis Facility
 distributed by the National Optical Astronomy Observatories which is operated
 by the Association of Universities for research in Astronomy, Inc. under
 co-operative agreement with the National Science Foundation} and photometry was
 performed using the stand-alone version of {\it DAOPHOT}\footnote{{\it DAOPHOT}
 stands for Dominion Astrophysical Observatory Photometry}
 \citep{1987PASP...99..191S}. 

 The SN was surrounded by a star forming knot and the host galaxy is highly
 inclined ($\sim82.3\degree$) with respect to the line of sight. The SN flux
 was, therefore highly contaminated by the background. In order to account for
 this, late-time ($\sim$ 36 months) high SNR images of the host galaxy in
 $UBVRI$ bands, obtained from the NOT under good photometric conditions, were
 used to map the galaxy flux, and a template subtraction technique
 \citep{2011ApJ...736...76R, 2011MNRAS.414..167R} was adopted to remove the
 galaxy contributions. Before flux subtraction, both NOT and ST images were
 brought to a common `plate-scale' using the `magnify' task provided in the
 $IRAF$ package. The instrumental magnitudes of the SN were derived from the
 template subtracted SN frames using the PSF fitting
 method.

 The field of \sn\ was calibrated from ST in the $BVRI$ bands using
 \citet{1992AJ....104..340L} standard stars of the fields of PG1633+009
 and PG1047+003 observed on the night of March 02, 2008 under moderate seeing
 (FWHM $\sim$ 2\arcsec\ in R band) and transparent sky conditions. We used mean
 values of the atmospheric extinction coefficients of the site 
 \citep[namely 0.28, 0.17, 0.11 and 0.07 mag per unit airmass for the $B$, $V$,
 $R$ and $I$ bands, from][]
 {2000BASI...28..675K} with typical standard deviations between the
 transformed and the standard magnitudes of Landolt stars of 0.04 in $B$, 0.02
 in $V$ and $R$ and 0.01 in $I$ band. A sample of 10 bright and isolated
 non-variable stars in the field of \sn\ were used as local standards to derive
 the zero points for the SN at each epoch.
  \begin{table*}
  \caption{{\it swift}/UVOT photometry of SN 2007uy.\label{tab:uvot_07uy}}
  \begin{tabular}{ccccccccc}
  \hline\hline
   UT Date  &   JD    &Phase$^{a}$&uvw2&uvm2&uvw1&u&b&v\\
  (yy/mm/dd)& 2454000+& (day)& (mag)& (mag)& (mag)& (mag)& (mag)& (mag)\\
  \hline
2008/01/06.15&471.65 & +9 &      21.45             &      24.33              & 18.61$\pm$0.18          &16.89$\pm$0.06     &     17.28$\pm$0.06 &     16.62$\pm$0.08\\
2008/01/06.49&471.99 &+10 &20.29$\pm$0.39          & 21.31$\pm$4.25          & 18.45$\pm$0.10          &16.93$\pm$0.04     &     17.02$\pm$0.03 &     16.47$\pm$0.04\\
2008/01/09.71&475.21 &+13 &20.05$\pm$0.35          & 20.15$\pm$0.58          & 18.23$\pm$0.10          &16.78$\pm$0.04     &     16.81$\pm$0.03 &     16.09$\pm$0.04\\
2008/01/11.38&476.88 &+14 &20.33$\pm$0.27          & 20.69$\pm$0.65          & 18.37$\pm$0.06          &16.69$\pm$0.02     &     16.79$\pm$0.02 &     15.96$\pm$0.02\\
2008/01/11.99&477.49 &+15 &21.28$\pm$0.82          &     $\textgreater$20.57 & 18.51$\pm$0.07          &16.72$\pm$0.02     &     16.81$\pm$0.02 &     15.93$\pm$0.02\\
2008/01/12.57&478.07 &+15 &20.48$\pm$0.40          & 21.40$\pm$3.72          & 18.45$\pm$0.09          &16.76$\pm$0.03     &     16.77$\pm$0.02 &     15.85$\pm$0.03\\
2008/01/13.21&478.71 &+16 &20.59$\pm$0.47          &      22.63              & 18.50$\pm$0.09          &16.76$\pm$0.03     &     16.80$\pm$0.03 &     15.93$\pm$0.03\\
2008/01/13.89&479.39 &+17 &20.34$\pm$0.29          & 20.81$\pm$0.68          & 18.41$\pm$0.07          &16.78$\pm$0.03     &     16.81$\pm$0.02 &     15.81$\pm$0.02\\
2008/01/14.73&480.23 &+18 &21.41$\pm$0.87          &     $\textgreater$20.59 & 18.61$\pm$0.08          &16.88$\pm$0.03     &     16.77$\pm$0.02 &     15.80$\pm$0.02\\
2008/01/15.49&480.99 &+18 &21.32$\pm$1.00          &     $\textgreater$20.36 & 18.95$\pm$0.12          &17.03$\pm$0.04     &     16.81$\pm$0.02 &     15.79$\pm$0.02\\
2008/01/16.29&481.79 &+19 &20.77$\pm$0.47          &     $\textgreater$20.36 & 18.85$\pm$0.11          &17.04$\pm$0.04     &     16.81$\pm$0.02 &     15.78$\pm$0.02\\
2008/01/17.14&482.64 &+20 &     21.43              &     $\textgreater$19.88 & 18.93$\pm$0.09          &17.15$\pm$0.03     &     16.90$\pm$0.03 &     15.77$\pm$0.02\\
2008/01/17.81&483.31 &+21 &21.53$\pm$0.96          &      22.97              & 18.94$\pm$0.08          &17.27$\pm$0.03     &     16.89$\pm$0.02 &     15.73$\pm$0.02\\
2008/01/18.85&484.35 &+22 &     $\textgreater$20.87&      22.19              &      $ - $              &     $ - $         &          $ - $     &          $ - $    \\
2008/01/21.12&486.62 &+23 &     22.92              &      24.98              & 19.62$\pm$0.11          &18.07$\pm$0.05     &     17.26$\pm$0.03 &     15.89$\pm$0.02\\
2008/01/21.76&487.26 &+25 &     $ - $              &      $ - $              & 19.71$\pm$0.40          &18.30$\pm$0.17     &     17.25$\pm$0.07 &     15.94$\pm$0.07\\
2008/01/25.17&490.67 &+28 &     $ - $              &      $ - $              &      $\textgreater$20.64&     $ - $         &          $ - $     &          $ - $    \\
2008/01/26.41&491.91 &+29 &     $ - $              &      $ - $              & 20.84$\pm$0.29          &19.33$\pm$0.10     &     17.88$\pm$0.03 &     16.28$\pm$0.02\\
2008/01/27.98&493.48 &+31 &     $ - $              &      $ - $              &      $\textgreater$20.91&     $ - $         &          $ - $     &          $ - $    \\
2008/01/29.35&494.85 &+32 &     $ - $              &      $ - $              &      $\textgreater$20.78&     $ - $         &          $ - $     &          $ - $    \\
2008/01/30.81&496.31 &+35 &     $ - $              &      $ - $              & 22.04$\pm$0.50          &20.78$\pm$0.58     &     18.51$\pm$0.06 &     16.66$\pm$0.03\\ 
2008/02/01.93&498.43 &+37 &     $ - $              &      $ - $              &      $\textgreater$20.87&21.59$\pm$1.57     &     18.53$\pm$0.06 &     16.86$\pm$0.03\\
2008/02/03.93&500.43 &+39 &     $ - $              &      $ - $              & 19.90$\pm$1.25          &     $ - $         &          $ - $     &          $ - $    \\
2008/02/08.95&505.45 &+43 &     $ - $              &      $ - $              & 22.39$\pm$0.50          &     $ - $         &          $ - $     &          $ - $    \\
2008/02/10.89&507.39 &+45 &     $ - $              &      $ - $              &      $ - $              &     $ - $         &     19.05$\pm$0.10 &          $ - $    \\
2008/02/13.03&509.53 &+47 &     $ - $              &      $ - $              &      $ - $              &     $ - $         &     18.95$\pm$0.09 &     17.27$\pm$0.05\\
2008/02/16.11&512.61 &+50 &     $ - $              &      $ - $              &      $ - $              &     $ - $         &          $ - $     &          $ - $    \\
2008/02/22.30&518.80 &+57 &     $ - $              &      $ - $              &      $ - $              &     $ - $         &          $ - $     &          $ - $    \\
2008/02/25.79&522.29 &+60 &     $ - $              &      $ - $              &      $ - $              &     $ - $         &     18.96$\pm$1.04 &     17.72$\pm$0.09\\
2008/03/12.60&538.10 &+75 &     $ - $              &      $ - $              &      $ - $              &     $ - $         &          $ - $     &     17.84$\pm$0.12\\
  \hline
  \end{tabular}
 \newline\newline
  $^{a}${With reference to the epoch of explosion JD 2454462.17}
  \end{table*}

\begin{table*}
\caption{Photometry of local standard stars in the field of SN 2007uy$^a$}
\label{tab:std_2007uy}
\begin{tabular}{ccccccc}
\hline\hline
Star & $\alpha_{2000}$ & $\delta_{2000}$ & $B$ & $V$ & $R$ & $I$\\
ID& (h m s)&(deg m s)&(mag)&(mag)&(mag)&(mag)\\   
\hline
1&09:09:42.48 &33:05:07.1&16.37$\pm$0.04&15.56$\pm$0.02&15.06$\pm$0.02&14.64$\pm$0.01\\
2&09:09:44.25 &33:05:11.8&17.64$\pm$0.04&16.98$\pm$0.02&16.62$\pm$0.03&16.27$\pm$0.01\\
3&09:09:37.26 &33:09:34.9&17.85$\pm$0.04&17.34$\pm$0.03&17.01$\pm$0.03&16.69$\pm$0.02\\
4&09:09:36.84 &33:05:46.0&18.00$\pm$0.05&17.26$\pm$0.02&16.84$\pm$0.03&16.43$\pm$0.01\\
5&09:09:36.12 &33:05:42.9&18.11$\pm$0.05&17.48$\pm$0.03&17.13$\pm$0.03&16.81$\pm$0.02\\
6&09:09:31.93 &33:05:19.5&19.01$\pm$0.05&17.93$\pm$0.04&17.26$\pm$0.04&16.74$\pm$0.01\\
7&09:09:18.95 &33:08:33.4&17.54$\pm$0.04&17.15$\pm$0.02&16.87$\pm$0.03&16.57$\pm$0.01\\
8&09:09:14.29 &33:07:16.0&18.85$\pm$0.05&17.49$\pm$0.03&16.63$\pm$0.03&15.92$\pm$0.01\\
9&09:09:11.82 &33:06:34.6&19.09$\pm$0.05&18.02$\pm$0.04&17.36$\pm$0.04&16.86$\pm$0.01\\
10&09:09:10.49&33:06:25.6&19.83$\pm$0.05&18.72$\pm$0.04&17.97$\pm$0.04&17.37$\pm$0.02\\
\hline
\end{tabular}
\newline\newline
 $^a$The $U$ band data is standardized with respect to the field
 standards mentioned in \citet{2009ApJ...692L..84M}. Errors are
 1$\sigma$ uncertainties and it include both photometric as well as
 calibration errors.
\end{table*}

 The location and magnitudes of these local standards are listed in Table
 \ref{tab:std_2007uy}. These secondary standards are marked in figure
 \ref{fig:snid}. For $U$ band, the calibration of \citet{2009ApJ...692L..84M} is
 used. Table \ref{tab:mag_2007uy} lists the final photometry of \sn\ in the
 $UBVRI$ bands at 37 phases between +16d and +129d.

\begin{table*}
\caption{$UBVRI$ photometry of SN 2007uy.}
\label{tab:mag_2007uy}
\begin{tabular}{ccccccccc}
\hline\hline
 UT Date & JD & Phase$^{a}$ & $U$ & $B$ & $V$ & $R$ & $I$ & Telescope$^{b}$\\
  (yyyy/mm/dd)& 2454000+& (day)& (mag)& (mag)& (mag)& (mag)& (mag)&\\
\hline
2008/01/12.94& 478.44& +16& 16.64$\pm$0.01& 16.78$\pm$0.01& 15.79$\pm$0.01& 15.54$\pm$0.01&      $ - $    & NOT\\
2008/01/13.99& 479.71& +17&      $ - $    & 16.77$\pm$0.02& 15.81$\pm$0.03& 15.51$\pm$0.03& 15.53$\pm$0.03& ST\\
2008/01/14.85& 480.58& +18&      $ - $    & 16.74$\pm$0.03& 15.78$\pm$0.03& 15.46$\pm$0.02& 15.43$\pm$0.04& ST\\
2008/01/15.19& 480.69& +18& 16.69$\pm$0.03& 16.71$\pm$0.02& 15.76$\pm$0.01& 15.38$\pm$0.02& 15.31$\pm$0.01& NOT\\
2008/01/15.99& 481.72& +19&      $ - $    & 16.78$\pm$0.02& 15.74$\pm$0.02& 15.44$\pm$0.01& 15.36$\pm$0.03& ST\\
2008/01/17.19& 482.69& +20& 16.95$\pm$0.01& 16.80$\pm$0.01& 15.81$\pm$0.01& 15.40$\pm$0.01& 15.21$\pm$0.01& NOT\\
2008/01/28.94& 494.44& +32&      $ - $    & 17.98$\pm$0.01& 16.45$\pm$0.01& 15.72$\pm$0.01& 15.50$\pm$0.01& NOT\\
2008/01/29.84& 495.57& +33&      $ - $    & 18.19$\pm$0.02& 16.44$\pm$0.02& 15.81$\pm$0.02& 15.59$\pm$0.02& ST\\
2008/01/30.19& 495.69& +33&      $ - $    &      $ - $    & 16.53$\pm$0.02& 15.79$\pm$0.01& 15.58$\pm$0.01& NOT\\
2008/01/31.94& 497.44& +35& 19.08$\pm$0.02& 18.22$\pm$0.01& 16.60$\pm$0.01& 15.89$\pm$0.01& 15.62$\pm$0.02& NOT\\
2008/01/31.87& 497.60& +35&      $ - $    & 18.28$\pm$0.03& 16.54$\pm$0.03& 15.91$\pm$0.03& 15.62$\pm$0.04& ST\\
2008/02/01.94& 498.44& +36& 19.20$\pm$0.03& 18.29$\pm$0.02& 16.66$\pm$0.02& 15.96$\pm$0.02& 15.66$\pm$0.02& NOT\\
2008/02/03.94& 500.44& +38& 19.32$\pm$0.03& 18.41$\pm$0.03& 16.86$\pm$0.02& 16.08$\pm$0.02& 15.77$\pm$0.02& NOT\\
2008/02/04.88& 501.61& +39&      $ - $    & 18.50$\pm$0.06& 16.82$\pm$0.02& 16.11$\pm$0.02& 15.83$\pm$0.03& ST\\
2008/02/09.75& 506.48& +44&      $ - $    & 18.70$\pm$0.03& 17.07$\pm$0.03& 16.40$\pm$0.03& 15.97$\pm$0.04& ST\\
2008/02/05.86& 512.59& +50&      $ - $    & 18.84$\pm$0.03& 17.33$\pm$0.01&      $ - $    & 16.18$\pm$0.03& ST\\
2008/02/18.94& 515.44& +53& 19.75$\pm$0.04& 19.01$\pm$0.03& 17.33$\pm$0.03& 16.66$\pm$0.03& 16.26$\pm$0.03& NOT\\
2008/02/23.65& 520.37& +57&      $ - $    & 19.27$\pm$0.05& 17.64$\pm$0.04& 16.84$\pm$0.03& 16.53$\pm$0.07& ST\\
2008/02/24.62& 521.35& +58&      $ - $    &      $ - $    & 17.52$\pm$0.02& 16.88$\pm$0.02& 16.44$\pm$0.03& ST\\
2008/02/29.94& 526.44& +64&      $ - $    &      $ - $    & 17.58$\pm$0.03&      $ - $    & 16.48$\pm$0.03& NOT\\
2008/03/02.75& 529.45& +66&      $ - $    & 19.22$\pm$0.07&      $ - $    &      $ - $    &      $ - $    & ST\\
2008/03/03.72& 531.53& +67&      $ - $    & 19.28$\pm$0.06& 17.66$\pm$0.04& 17.05$\pm$0.03& 16.54$\pm$0.05& ST\\
2008/03/05.80& 533.61& +69&      $ - $    & 19.29$\pm$0.03&      $ - $    & 17.05$\pm$0.04& 16.53$\pm$0.06& ST\\
2008/03/17.69& 543.19& +81&      $ - $    & 19.31$\pm$0.04& 17.81$\pm$0.03& 17.29$\pm$0.03& 16.76$\pm$0.03& NOT\\
2008/03/05.80& 587.36&+124&      $ - $    &      $ - $    & 18.67$\pm$0.02& 18.32$\pm$0.03& 18.02$\pm$0.06& ST\\
2008/05/02.61& 589.34&+126&      $ - $    &      $ - $    &      $ - $    & 18.51$\pm$0.05&      $ - $    & ST\\
2008/05/05.62& 592.35&+129&      $ - $    &      $ - $    & 18.84$\pm$0.03&      $ - $    &      $ - $    & ST\\
\hline
\end{tabular}
\newline\newline
$^{a}$With reference to the epoch of explosion JD 2454462.17.\\
$^{b}$The photometric observations are taken with the 1-m 
Sampurnanand Telescope (ST), ARIES, Nainital and 2.6-m Nordic Optical 
Telescope (NOT) with ALFOSC detector. Errors in magnitude denote $1\sigma$ 
 uncertainty.
\end{table*}

 The field was also monitored in $JHK$ NIR bands at 8 phases between
 +16d and +87d using United Kingdom Infrared Telescope (UKIRT) with Wide-Field
 Camera (WFCAM) as a backend detector \citep{2006SPIE.6269E..31H}. The
 dithered images were processed with standard tasks in $IRAF$ and photometry was
\begin{table*}
\caption{Near-infrared $JHK$ photometry of SN 2007uy.}
\label{tab:ukirt_2007uy}
\begin{tabular}{cccccc}
\hline\hline
 {UT Date}& {JD}& {Phase$^{a}$}& {$J$}& {$H$}& {$K$}\\
 (yyyy/mm/dd)& 2454000+& (day)& (mag)& (mag)& (mag)\\
\hline
2008/01/12.67& 478.17&+16& 14.48$\pm$0.25& 14.98$\pm$0.25&  14.98$\pm$0.17\\
2008/01/14.68& 480.18&+18& 14.32$\pm$0.19& 14.80$\pm$0.24&  14.71$\pm$0.14\\
2008/01/15.66& 481.16&+19& 14.26$\pm$0.19& 14.73$\pm$0.24&  14.67$\pm$0.15\\
2008/01/17.69& 483.19&+21& 14.12$\pm$0.32& 14.64$\pm$0.28&  14.51$\pm$0.22\\
2008/01/23.67& 489.17&+27& 14.13$\pm$0.25& 14.60$\pm$0.25&  14.51$\pm$0.25\\
2008/02/15.92& 512.42&+50& 15.76$\pm$0.07& 15.38$\pm$0.02&  15.06$\pm$0.18\\
2008/02/24.92& 521.42&+59& 16.07$\pm$0.06& 15.58$\pm$0.03&  15.32$\pm$0.19\\
2008/03/23.92& 549.42&+87& 17.03$\pm$0.07& 16.36$\pm$0.03&  16.21$\pm$0.19\\
\hline
\end{tabular}
\newline\newline
$^{a}${With reference to the epoch of explosion JD 2454462.17.}
\end{table*}

 performed with $DAOPHOT$ routines. The field was calibrated with respect
 to nearby 2MASS stars, and the calibrated $JHK$ magnitudes of \sn\ is given in
 Table \ref{tab:ukirt_2007uy}.

\subsection{Radio Data}
\label{sec:xray}
 The transient was first detected in radio on 2008 January 6.18 UT, using the
 Very Large Array (VLA) at 8.46 GHz, with a flux density of
 {290}$\pm$30\,$\mu$Jy \citep{2008ATel.1350....1S}. Although the VLA
 observations originally targeted SN 2007uy, after January 10.2 UT the field of
 observation was centred on another bright transient, SN 2008D, discovered in
 the same galaxy. Due to larger primary beam size of VLA in L (30$\arcmin$), C
 (9$\arcmin$) and X ($5.\arcmin4$) bands, the locations of both transients were
 within VLA field of view. Here, L band corresponds to 1.34-1.73 GHz, C band
 corresponds to 4.5-5.0 GHz and X band corresponds to 8.0-8.8 GHz radiation. We
 used the NRAO archival facility to fetch these data sets and reduced the data
 using standard routines of
  \begin{table*}
  \caption{Log of radio observation of SN 2007uy from VLA in 4.8 \& 8.4 GHz.}
  \centering
  \label{tab:vladata}
  \begin{tabular}{cccccc}
  \hline\hline
   UT Date&JD&Phase$^a$&Frequency&Flux&Flux error$^b$\\
  (yy/mm/dd)& 2454000+& (day)& (GHz)& (mJy)& (mJy)\\
  \hline
  2008/01/06&471.76& +09&8.4&0.362&0.044\\
  2008/01/07&473.11& +10&8.4&0.306&0.070\\
  2008/01/11&476.82& +14&8.4&0.459&0.469\\
  2008/01/14&479.76& +17&8.4&0.513&0.040\\
  2008/01/16&481.80& +19&4.8&0.515&0.050\\
  2008/01/16&481.83& +19&8.4&0.688&0.057\\
  2008/01/17&482.90& +20&4.8&0.403&0.037\\
  2008/01/17&482.92& +20&8.4&0.566&0.053\\
  2008/01/20&485.66& +23&4.8&0.562&0.044\\
  2008/01/21&486.66& +24&8.4&0.724&0.060\\
  2008/01/21&486.67& +24&4.8&0.799&0.055\\
  2008/01/23&488.66& +26&4.8&0.770&0.060\\
  2008/01/23&488.67& +26&8.4&0.721&0.076\\
  2008/01/25&490.67& +28&8.4&0.797&0.056\\
  2008/01/25&490.68& +28&4.8&0.875&0.068\\
  2008/01/27&492.92& +30&8.4&0.965&0.081\\
  2008/01/27&492.93& +30&4.8&0.821&0.050\\
  2008/01/30&495.82& +33&8.4&0.953&0.070\\
  2008/02/01&497.73& +35&4.8&1.216&0.085\\
  2008/02/01&497.74& +35&8.4&0.968&0.068\\
  2008/02/03&499.70& +37&8.4&1.044&0.075\\
  2008/02/03&499.71& +37&4.8&1.072&0.065\\
  2008/02/08&504.72& +42&4.8&1.263&0.078\\
  2008/02/08&504.73& +42&8.4&1.011&0.075\\
  2008/02/14&510.79& +48&8.4&0.872&0.063\\
  2008/02/14&510.80& +48&4.8&1.249&0.085\\
  2008/02/21&517.92& +55&4.8&1.276&0.102\\
  2008/02/21&517.93& +55&8.4&0.522&0.048\\
  2008/02/24&520.65& +58&4.8&1.432&0.082\\
  2008/02/24&520.66& +58&8.4&0.677&0.046\\
  2008/03/07&532.60& +70&4.8&1.339&0.106\\
  2008/03/07&532.62& +70&8.4&0.596&0.062\\
  2008/03/21&546.58& +84&4.8&1.103&0.089\\
  2008/03/22&547.65& +85&8.4&0.351&0.049\\
  2008/04/14&570.53&+108&4.8&0.814&0.120\\
  2008/05/04&591.36&+129&4.8&0.457&0.107\\
  \hline
  \end{tabular}
\newline\newline
$^{a}${With reference to the explosion epoch JD 2454466.17}\\
$^b${The Flux errors are measured using the expression ${\sigma}_f^2 = {(\epsilon.S_0)}^2 + {\sigma}_0^2 + {\sigma}_{S_0}^2$. Here $S_0$ is the observed flux density, $\sigma_0$ is the RMS noise of the radio sky, ${\sigma}_{S_0}$ is the error
 associated with $\sigma_0$ and $\epsilon$ is the fraction that accounts the error in VLA flux calibration. For 4.8 and 8.4 GHz observations value of $\epsilon$ is 0.05. For the epochs where measured flux density is less than 3 times of the corresponding value of ${\sigma}_f$, we have considered the flux density as the
 upper limit for our measurement.}
  \end{table*}

 `Astronomical Image Processing System' (AIPS)\footnote{
 Astronomical Image Processing System (AIPS) has been developed by the National
 Radio Astronomical Observatories (NRAO), USA}. The transient was not detected
 in the radio U (14.4-15.4 GHz) band, but was prominent in the C and X band
 images. The new VLA
 data set is presented in Table \ref{tab:vladata}. For further analysis we have
 used the literature data \citep{2011ApJ...726...99V} along with the VLA
 archival data.

\subsection{Low-resolution optical spectroscopy}
\label{sec:spec}
 The spectroscopic observations of \sn\ were carried out at 7 epochs between +17d
 to +392d. The spectral data for +17d were acquired with NOT/ALFOSC on January
 13, 2008. The spectra for the epochs +32d, +58d, +122d \& +392d are based on
 the archival data obtained through `ESO Science Archive Portal' which were
 acquired using 8m VLT and 3.6m NTT. The spectral data for +96d and
 +162d were taken from \citet{2010ApJ...709.1343M}. These were primarily
 acquired on 1st April and 6th June, 2008 using the `MMTBLUECHANNEL' detector at
 MMT. A journal of spectroscopic observations is presented in Table
 \ref{tab:speclog}.

 All the raw optical data were processed using the standard tasks in $IRAF$.
 Bias and flat-fielding were performed on each frame. Cosmic ray rejection on
 each frame was done by using  Laplacian kernel detection routine LACOSMIC
 \citep{2001PASP..113.1420V}\footnote{http://www.astro.yale.edu/dokkum/lacosmic}.
\begin{figure*}
\centering
\includegraphics[width=17cm]{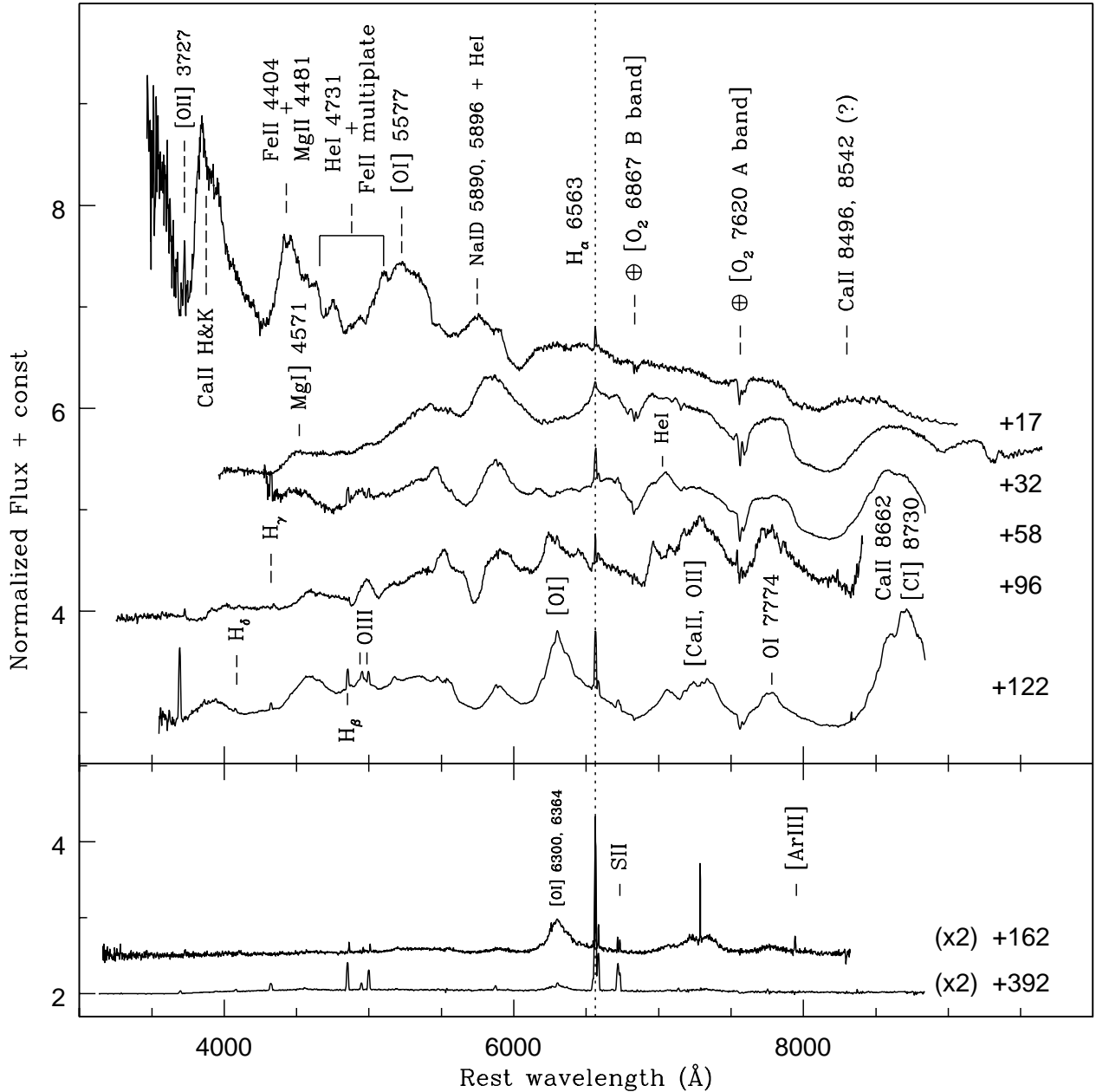}%
\caption{Spectroscopic evolution of SN 2007uy.
 All the spectra have been normalized with respect to the peak flux of the
 underling \ha\ feature and a constant offset has been applied to present them
 clearly. The +162d and +392d spectra have been multiplied by a factor of 2 to
 enlarge several tiny features. The dotted vertical line represents the position
 of \ha\, and confirms the wavelength calibration within the limits of the
 spectral resolution.}
\label{fig:spectra}
\end{figure*}

\begin{table*}
\caption{Journal of spectroscopic observations of \sn\,$^{a}$.}
\label{tab:speclog}
\begin{tabular}{l cc cc cc cc}
\hline\hline
 {UT Date}& {Phase$^{b}$}& {Range}& {Telescope$^{c}$}&
 {Grating}& {Slit width}& {Dispersion}& {Exposure}& {SNR$^{d}$} \\
 (yy/mm/dd)& (days)& \mum&     & (gr mm$^{-1}$)&  (\arcsec)& (\AA\,pix$^{-1}$)&  (s)&(pix$^{-1}$)\\
\hline
        2008/01/13& +17 & 0.32$-$0.91& NOT& 300& 1.3 & 3.0& 1200&135\\
        2008/01/28& +32 & 0.38$-$0.92& NTT& 300& 1.0 & 1.7& 1200&170\\
        2008/02/22& +58 & 0.45$-$0.87& VLT& 300& 1.3 & 1.7&  900&225\\
        2008/04/01& +96 & 0.32$-$0.80& MMT& 300& 1.0 & 2.0&  600$\times$4&110\\
        2008/04/26&+122 & 0.35$-$0.87& VLT& 300& 1.3 & 1.7& 2940&250\\
        2008/06/06&+162 & 0.38$-$0.68& MMT& 300& 1.0 & 2.0&  900&80\\
        2009/01/20&+392 & 0.33$-$0.87& VLT& 300& 1.3 & 1.7& 3600&50\\
\hline
\end{tabular}
\newline\newline
$^{a}${The spectra for +32, +58, +122 \& +392d are based on data
 obtained from the ESO Science Archive Facility. The spectra for +96 \& +162d
 are taken from \citet{2010ApJ...709.1343M}.}\\
$^{b}${With reference to the epoch of explosion JD 2454462.17.}\\
$^{c}${NOT : ALFOSC on 2.6-m Nordic Optical Telescope (NOT), la
   Palma; NTT : EMMI on 3.6-m New Technology Telescope, la Silla; VLT : FORS2
  on 8.2-m ESO-VLT-UT1 telescope, Paranal Observatory; MMT : Blue Chanel
 spectrograph on 6.5-m MMT at Mt. Hopkins; }\\
$^{d}${Signal-to-noise-ratio at 0.6\mum.}
\end{table*}

 Images were co-added to improve the signal-to-noise ratio and one-dimensional
 spectra were extracted from co-added frames using the {\it apall} task in IRAF
 \citep{1986PASP...98..609H}. Wavelength calibration was performed using the
 {\it identify} task and
 the fifth order fits were used to achieve a typical RMS scatter of 0.1\AA
 (\ie\,, 60\kms\ at 5000\AA\,). The position of the \Oi\ emission skyline at
 5577\AA\, was used to check the wavelength calibration and deviations were
 found to lie between 0.5 to 1\AA\, and this was corrected by applying a linear
 shift in wavelength.
 
\section{Spectroscopic evolution}
\label{sec:specevo}
 In order to proceed further with the spectral analysis, the wavelength
 calibrated spectra were corrected for the recession velocity of the host galaxy
 using the prominent Balmer emission lines arising from the underlying \Hii\
 regions. In this way both the redshifts due to recession and rotation of the
 galaxy have been accounted for.
 All the spectra were normalized with respect to the peak flux of the
 underlying \ha\ feature and a constant offset was applied to present them
 clearly.
 In figure \ref{fig:spectra}, the spectra of \sn\ is presented at 7 epochs
 spanning between +17d and +392d. The +17d spectrum was taken near maximum
 light, while +162d and +392d are late nebular phase spectra. The latter
 are enlarged by a factor of 2 for clarity.
 Spectral features are mainly identified as per
 previously published line identification lists for Type Ibc SNe
 (\citealt{2000asqu.book.....C} and references therein), though high degree of
 line-blending and line-blanketing limit the detections. The dotted
 vertical line represents the position of \ha\,. We have marked few nebular
 lines like \ha\,, \Oiia\ $\lambda$3727, which belong to the associated star
 forming region and is appeared starting from the early spectrum.
 The absence of Silicon (Si) feature at 6315\AA\ as well as the absence of H
 lines indicate that the spectra of \sn\ resemble those of stripped-envelope
 SNe. The lines are highly blended with each other and for most of the lines,
 the P-Cygni profiles are affected by `line-blanketing'. The metallic lines like
 \Feii\ $\lambda$4401, \Mgii\ $\lambda$4481 as well as \Feiia\ $\lambda$5536,
 \Nai\,D $\lambda\lambda$5890, 5896 and \Caii\ $\lambda\lambda$8496, 8542 are
 prominent from the early epochs. In the +17d spectrum, the features
 between 4700\AA\ and 5000\AA\ are plausibly attributed by the blend of \Hei\
 $\lambda$4731 along with \Feii\ multiplates. The lines of O, Mg and Ca start
 to appear nearly
 $\sim$30d after the explosion. The spectra evolved faster than other Type Ibc
 SNe and except \Oia\ $\lambda\lambda$6300, 6364; \Caiia\ and \Oiia\,, all other
 features almost faded out by +162d. This is in contrast with normal Type
 Ibc SNe, where emission line profiles remain prominent even beyond +250d (see
 for e.g., \citealt{2000asqu.book.....C, 2008ApJ...687L...9M,
 2010ApJ...709.1343M, 2011MNRAS.416.3138V}).

 The spectral lines \Oia\ $\lambda5577$ and \Oia\ $\lambda\lambda6300,6364$ show
 highly blueshifted emission profiles in the early phase spectra, before
 eventually moving to their respective rest wavelengths at the +92d phase and
 onwards. The high blueshift of the spectral features is a signature of aspheric
 explosion. The blueshift of the spectral lines at early epochs
 was also observed in other Type I events, especially in the case of Type Ia
 events \citep{2010Natur.466...82M, 2010ApJ...708.1703M, 2011MNRAS.413.3075M}.

\begin{figure}
\centering
\includegraphics[width=8.5cm]{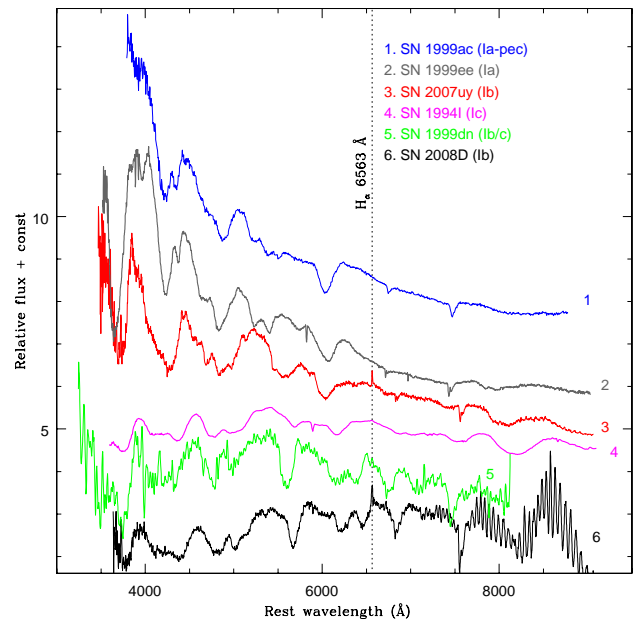}%
\caption{Spectroscopic comparison of \sn\ with other Type I events near maxima.
 The colour version of the figure is available in the online journal.}
\label{fig:spec_comp}
\end{figure}

 In figure \ref{fig:spec_comp} we compare the +17d spectrum of \sn\,  with
 the spectra of Type Ia, Ia-pec, Ibc and Ic events, taken at similar phases. The
 blueshift in the bluer part of the spectrum is clearly visible; and, the
 spectrum of \sn\ is highly similar to that of Type Ia and Ia-pec events (Maeda
 et al. 2010a,b; 2011). The characteristics that discriminate \sn\ from Type Ia
 events are: the presence of \Hei\ lines as well as strong \ha\ which indicates
 that the progenitor is associated with a star forming region, a common site
 for Type Ibc events. Moreover \sn\ is radio luminous (\S\ref{sec:xray}), which
 is the result of interaction of the SN shock with the circumburst medium. Type
 Ia SNe progenitors are not surrounded by a dense circumstellar medium and hence
 are very rarely observable in the radio in the early epochs
 \citep{2012ApJ...750..164C, 2002ARA&A..40..387W}.

\subsection{Evolution of some spectral lines: a message regarding the
 aspheric explosion}
\label{sec:asym}
\begin{figure*}
\centering
\includegraphics[width=17cm]{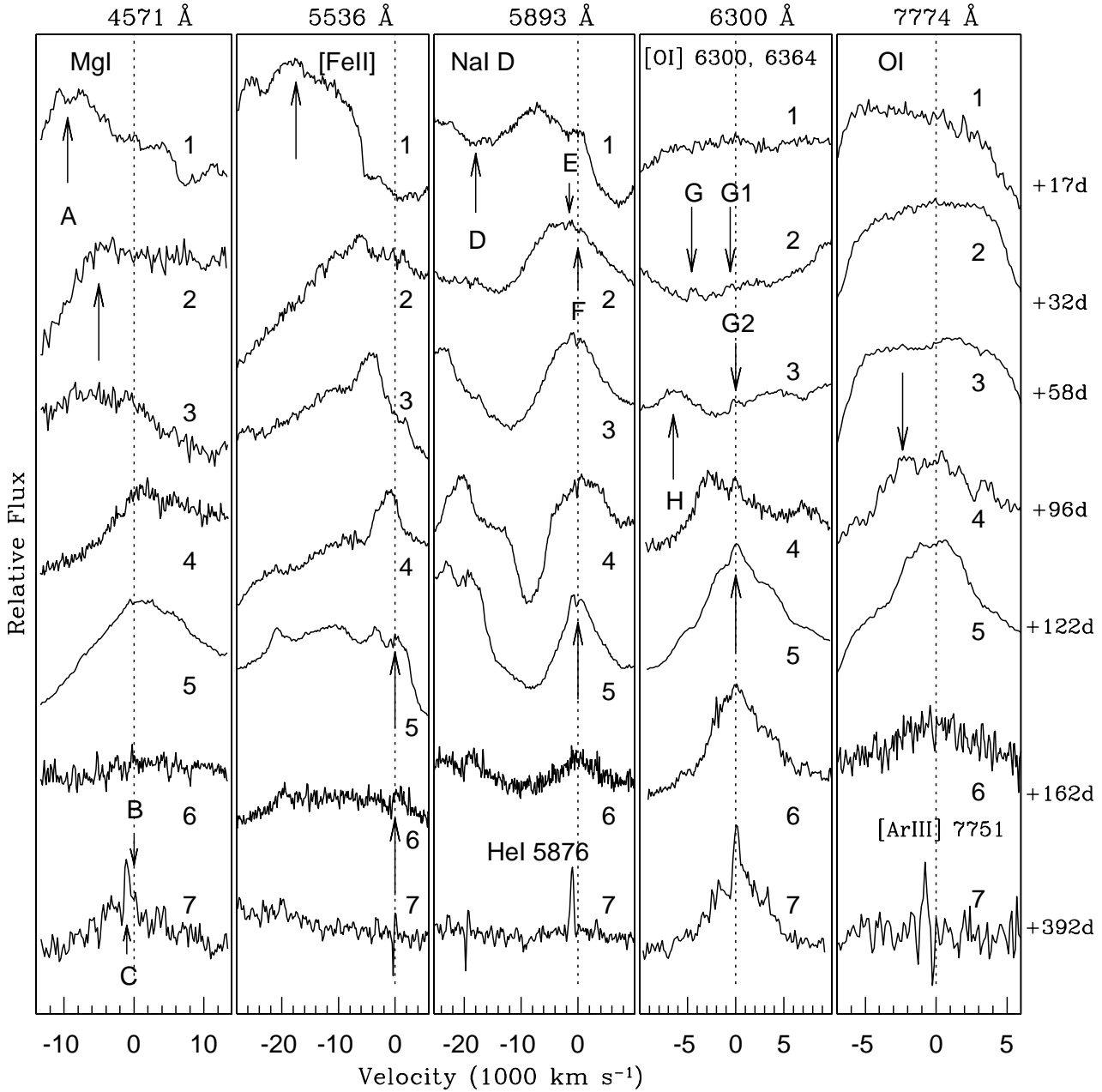}%
\caption{Temporal evolution of some spectral lines of \sn\,. The zero velocity
 shown with dotted line in each panel refers the rest wavelength
 of corresponding elements as mentioned at the top of the panels. The flux scale
 is relative. To make the features prominent, the intensity of each feature has
 been normalized by the flux of the zeroth velocity of the corresponding
 feature, though all the flux related measurements have been done using the
 normalized spectra shown in figure \ref{fig:spectra}.}
\label{fig:linevel}
\end{figure*}

 Figure \ref{fig:linevel} presents the spectral evolution of \Mgia\,,
 $\lambda$4571, \Feiia\ $\lambda$5536, \Nai\,D $\lambda\lambda$5890, 5896,
 \Oia\ $\lambda\lambda$6300, 6364 and \Oi\ $\lambda$7774, which are commonly
 found in Type Ibc events and in Type IIP SNe.
 During the early epochs
 most of the features are highly blueshifted with respect to their rest
 wavelengths / velocities.


 The early and late phase blue-shifting of the spectral lines can arise for
 different reasons, \eg\,, (i) contamination due to other emission lines
 \citep{2003MNRAS.338..939E}, (ii) a torus or elongated ejecta with a
 sufficiently opaque inner portion \citep{1992SvAL...18..168C,
 1994Natur.369..380W} or (iii) due to dust formation \citep{1989IAUC.4746....1D,
 1989LNP...350..164L, 2003MNRAS.338..939E}. For the formation of dust which can
 block the red wings of the spectral lines, the presence of a cold SN shell is
 required, whereas newly formed dust grains can only increase the continuum part
 (as discussed in \S\ref{sec:secondepoch}). With a Monte-Carlo simulation
 \citet{2009MNRAS.397..677T} showed that, due to the residual opacity of the
 core ejecta, the emission lines can be blueshifted only about by $30-40$ \AA\,,
 which is much less than the blueshifts ($\sim 70-$100\AA\,) observed for \sn\,.
 Thus, the third possibility can be ruled out for young ($\la$100d) hot SN
 ejecta.

 Asphericity in the SN explosion, with a torus or disk-like distribution of
 different elements perpendicular to a semi bipolar jet, could also be an
 explanation for
 the aspheric signature of the lines viz, double-horn emission profile of \Oia\
 $\lambda\lambda$6300, 6364 (\citealt{2008Sci...319.1220M, 2008ApJ...687L...9M}
 and references therein). The residual opacity in the core of elongated
 ejecta may also be responsible for the observed blueshift. Optically thick
 inner ejecta could obstruct the light from the rear side of the SN, creating a
 flux deficit in the redshifted part of the emission lines
 \citep{1992SvAL...18..168C, 1994Natur.369..380W}. It is worth mentioning that
 symmetric double-peaked profiles can also be
 explained by the doublet nature of \Oia\ $\lambda\lambda$6300, 6364 seen
 under optically thick conditions leading to an intensity ratio close to one
 \citep{2010ApJ...709.1343M}. Here we discuss the spectral evolution of each
 feature.

 The \Mgia\,$\lambda$4571 (the first panel from the left of figure
 \ref{fig:linevel}) is seen clearly in +35d spectrum as a broad (FWHM
 $\sim$185\AA), highly blue-shifted 
 ($\sim$71\AA) emission peak that corresponds to $\sim$4660 km~s$^{-1}$
 blueshifted velocity of the emitting region projected onto the line of sight.
 The `double-horn' feature marked with {\tt A} at the extreme blue end, is also
 seen in +17d spectrum. The \Mgia\, feature starts to move towards its
 rest wavelength after its first appearance at +32d and remains as a broad
 emission peak (FWHM $\sim$205\AA) until +122d with the emission peak at zero
 velocity. This broad feature almost disappears in the +162d spectrum
 and remains as a weak feature marked as {\tt B} in the deep nebular spectrum
 obtained at +392d. The feature {\tt B} is actually blended with a relatively
 blue-shifted broad emission feature {\tt C}. After deblending these two
 features, we marked the feature {\tt C} as \Baii\ $\lambda$4554
 \citep{2000asqu.book.....C}.

 The appearance of the extremely blueshifted (velocity $\sim -$17500 
 km~s$^{-1}$) emission peak of the blended lines of \Oia\ $\lambda$5577, \Feiia\
 $\lambda$5536 and \Coiia\ $\lambda$5526 at +17d can be seen clearly (the second
 panel from the left of figure \ref{fig:linevel}). A similar
 blueshifted emission was initially identified as \Oia\ $\lambda$5577 for SN
 1993J \citep{1994MNRAS.266L..61S, 1994AJ....108.2220F,
 1994Natur.369..380W} and for \sn\ \citep{2010ApJ...709.1343M}, although the
 study of \citet{1996ApJ...456..811H} showed that this emission is a blend of
 \Oia\,, \Feiia\ and \Coiia\,. Over time this feature developed and started
 to shift toward its rest wavelength from +122d onwards. This confirms the
 prediction of \citet{1996ApJ...456..811H} and establishes this emission line as
 a blend. Similar to \Mgia\,, \Feiia\ $\lambda$5536 also showed
 a symmetric emission profile at around zero velocity in late epoch ($\ga$122d)
 spectra. Interestingly from +122d (labeled {\tt 5} in the second panel,
 from left, in figure \ref{fig:linevel}) the intensity of this blended
 line dropped drastically. The blueshifted emission peak of \Oia\
 $\lambda5577$ in the late time evolution of the SN ejecta is commonly observed
 in Type Ibc SNe, \eg\ 2004ao, 2006T, 2008D and 2008bo
 \citep{2010ApJ...709.1343M}, but in none of the cases was a corresponding
 redshifted component was found. 

 The \Nai\,D feature (the third panel from the left of figure \ref{fig:linevel})
 is more or less a perfect P-Cygni profile, nonetheless it
 also shows a highly blueshifted (velocity $\sim$7700 
 km s$^{-1}$) emission peak, possibly blended with the \Hei\ $\lambda$5876
 emission feature. The corresponding \Hei\ absorption dip is marked as {\tt D}
 at +17d. The velocity drops down to about 1933 km s$^{-1}$ at +32d and the
 feature becomes almost like a P-Cygni profile from +58d. The \Nai\,D feature is
 absent in +392d spectrum and the \Hei\
 $\lambda$5876 emission feature of the underlying star forming region is clearly
 visible at its rest wavelength. A careful inspection shows the existence of a
 tiny emission peak {\tt E} and an absorption dip {\tt F} at `zero velocity',
 starting from the initial epoch (+32d). We speculate that {\tt E} arises from
 of \Hei\,, while {\tt F} is a footprint of intervening \Nai\,D absorption due
 to the host galaxy. There are two tiny absorption features around {\tt E} $-$ one
 is at the right {\tt F} and other at the left, about 2000 \kms\ blueshifted with respect to {\tt
 F} that is equal to the redshift of the host. Hence the possibility of \Nai\,D
 impressions due to host (feature {\tt F}) and Milky Way (blueshifted feature)
 can not be ruled out.

 The \Oia\ $\lambda\lambda$6300, 6364 is a frequently studied emission line in
 the context of the asymmetric nature of Type Ibc explosions. Different
 geometries of oxygen ejecta have been proposed as the origin of the observed
 line profiles of this doublet. A flat-topped profile is produced by a radially
 expanding spherical shell of oxygen gas; a parabolic profile is produced by a
 filled uniform sphere; and the double horned profile is due to the presence of
 a cylindrical ring or torus like structure, that expands in the equatorial
 plane along the line of sight, where the bulk
 of the emitting gas is located at the projected expansion velocity ($\pm v$) of
 the torus and a jet-like ejection of matter is along the direction
 perpendicular to the line of sight. The `$\pm$' sign represents the red and
 blueshifted projected velocity of the torus respectively
 (\citealt{2008Sci...319.1220M, 2008ApJ...687L...9M} and references therein).
 Alternatively it can be interpreted as the doublet nature of \Oia\
 $\lambda\lambda$6300, 6364, seen in an optically thick environment
 \citep{2010ApJ...709.1343M}. In the fourth panel of figure
 \ref{fig:linevel} we have shown its evolution. In +32d spectrum a tiny emission
 feature {\tt G} is appeared at a blueshifted velocity $\sim$5839 km s$^{-1}$.
 Apparently it
 seems that by +58d, this feature moves to a new position {\tt G2} along with
 the appearance of a second peak {\tt H} at the bluer end. However, due to lack
 of spectrum between +32d and +58d, this dynamics can not be supported strongly.
 The velocity of {\tt H} with respect to rest position of \Oia\ $\lambda$6300 is
 $\sim$7732 km s$^{-1}$. This corresponds to a wavelength separation of about
 163\AA.
 Undoubtedly {\tt G2} represents the rest wavelength of \Oia\ $\lambda$6300 and
 it has been found to be preserved during follow-up observations. {\tt G2} may
 also be an evolved stage of {\tt G1}, a very tiny feature
 just above the noise level in +32d spectrum. The feature {\tt G} at 6205 \AA\
 may be due to a blend of \Feii\ and \Caii\ multiplates along with \Nii\,,
 and \Tiii\,, whereas since {\tt G1} starts to appear from early epochs, it is
 possibly due to a shallow spherical distribution of \Oia\ at the outer ejecta.
 At +96d the separation between {\tt H} and {\tt G1} is reduced to about 64\AA,
 but in the +122d, +162d and +392d spectra the emission peak profile became
 almost symmetrical around `zero velocity'. This indicates an uniform spherical
 distribution of \Oia\ $\lambda\lambda$6300, 6364 from +122d onward. Since {\tt
 G1} and {\tt H} do not appear simultaneously, their internal separation is
 varying with time and the intensity ratio between the blue and red wings of
 the double horn feature does not tend toward 3:1 with time; we rule out the
 possibility of appearance of the doublet nature of \Oia\ $\lambda\lambda$6300,
 6364 in the form of a double horn profile, at least for this particular case.
 Since these features start to appear from a very early phase, the jet and
 disk/torus-like scenario, which is invoked to explain late time ($\ga$200 days)
 \Oia\ features, may not be applicable to explain this particular spectral
 evolution. Similarly the `Flat-top' scenario is also not applicable.
  
 Unlike blended \Oia\ $\lambda5577$ and \Oia\ $\lambda\lambda6300, 6364$, the
 \Oi\ $\lambda7774$ feature, shown in the right most panel of figure
 \ref{fig:linevel}, developed as a symmetrical flat-topped peak, centered around
 `zero velocity'. Although the present spectra do not cover the emergence and
 development of the \Oi\ $\lambda7774$ line explicitly, they give a clear indication
 regarding the development of this particular line at around +96d from a
 completely featureless part of the spectrum. At +122d it becomes prominent,
 though it disappears at +162d. The flat-topped peak is a clear indication of a
 shell-like structure of the line emitting region and a rapid dilution of the
 line emitting region is possibly responsible for the sudden decrease of line
 intensity. At +392d, \Oi\ $\lambda7774$ disappears completely, rather we notice
 the impression of \Ariii\ $\lambda7751$ at its rest wavelength, about 1000
 km\,s$^{-1}$ blueshifted with respect to \Oi\ $\lambda7774$ in velocity domain.

 From the above discussion it seems that the velocity profiles of different
 line forming regions are not identical for \sn\,  and hence probably the
 distribution of \nickel\ does not influence the distribution of different
 material inside the ejecta of \sn\,, otherwise the evolution of all material
 would be of similar. The progenitors of Type Ib SNe have mainly He and O rich
 shells \citep{1990ApJ...361L..23S}. From the geometry of line profiles
 \citep{2008ApJ...687L...9M}, we propose that \Oi\ $\lambda7774$ line forming
 region was distributed inside the ejecta in
 the form of a shell whereas \Nai\,D and Mg ions were concentrated toward the
 inner portion of the ejecta with a roughly spherical distribution. Evolution of
 the blended \Oiia\ $\lambda5577$ and \Oiia\ $\lambda6300$ lines are especially
 interesting. The projected velocity of {\tt H} show a power law profile
 with temporal decay index $-1.6$ where the decay indices of \Mgia\,, \Nai\,D
 and \Oiia\ $\lambda5577$ blend (rest wavelength $\sim5540$ \AA\,) are
 respectively $-1.5$, $-1.6$ and $-1.3$. This confirms that the regions which
 are forming the lines \Oiia\ $\lambda5577$ blend, feature {\tt H}, \Mgia\, and
 \Nai\,D are `attached' with each other.
 Interestingly, the
 \Oiia\ $\lambda5577$ blend shows as a very tiny feature (FWHM $\sim41$ \AA\,)
 at +32d, before the emergence of feature {\tt H} and became prominent (FWHM
 $\sim63-68$ \AA\,) in the +58d and +96d spectra during the appearance of
 feature {\tt H}, before again shrinking to a tiny feature with FWHM $\sim34$
 \AA\ at +122d when {\tt G2} and {\tt H} merged together.

\begin{figure}
\centering
\includegraphics[width=8.5cm]{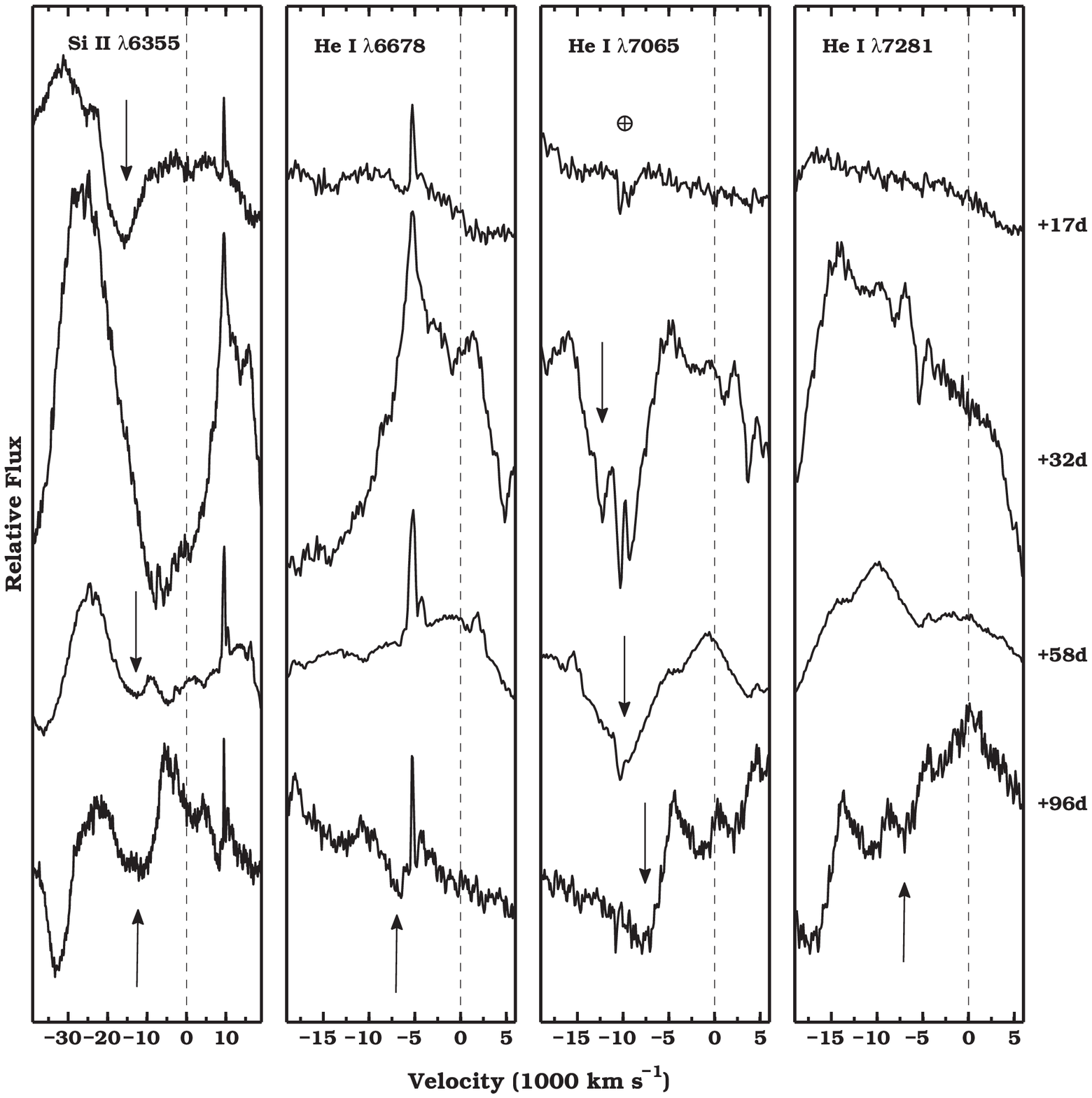}%
\vskip 0.5cm
\caption{Temporal evolution of velocities of SiII and HeI spectral features in
 \sn\,. }
\label{fig:linevel_si_HeI}
\end{figure}

 By +122d {\tt H} meets {\tt G2} (or
 both settle down to a constant velocity) and starts to move in a spherically
 symmetric manner and the ejecta becomes sufficiently cool. Our photometric
 calculation shows that temperature of this ejecta should asymptotically reach
 below 5000K (see \S\ref{sec:secondepoch}). In a simulation
 on forbidden lines in astrophysical-jets \citet{2004ApJ...609..261H} showed
 that below 5000K the strength of \Oia\ $\lambda 5577$ lessens
 in comparison to that of \Oia\ $\lambda 6300$ ($\la$0.1), even with a
 moderate electron number density (see figure 9 of
 \citealt{2004ApJ...609..261H}). Here we speculate that this could be the reason
 for the disappearance of \Oia\ $\lambda 5577$ line in between +96d and +122d.

 The spectral evolution at the locations of the most prominent \Hei\ lines is
 displayed in figure \ref{fig:linevel_si_HeI}. \Hei\ $\lambda\lambda$ 6678, 7065
 and 7281 are the strongest and, at the same time, most isolated He lines in the
 optical spectrum of a SN and are thus often used to distinguish between a SN Ib
 and Ic. In contrast, \Hei\ $\lambda$ 5876 is often blended with \Nai\,D
 (figure \ref{fig:linevel}), a line that also appears in SNe Ic, and \Hei\
 $\lambda$ 4471 is in a region of the spectrum with other overlapping
 transitions, including Fe lines. Our spectrum at maximum light (+17d) does not
 show any trace of He in any reasonable velocities. \Hei\ $\lambda$ 7065 appears
 15 days later at a velocity of $\sim$12000 \kms\ (although the line is blended
 with the telluric B-band). In the same spectrum we do not see any other
 credible He feature. The same holds for the spectrum at +58d, where \Hei\
 $\lambda$ 7065 has decelerated to 10000 \kms\,. \Hei\ $\lambda\lambda$ 6678,
 7281 finally appear in the +96d spectrum in a consistent velocity with \Hei\
 $\lambda$ 7065 ($\sim$7000 \kms\,). Figure \ref{fig:linevel_si_HeI} also shows
 the region around 6000 \AA\,. This region is dominated by an absorption feature
 that is usually attributed to \SiII\ $\lambda$ 6355 in most Type I SNe.
 Adopting this identification, this line decelerates from $\sim$15000 \kms\ at
 +17d to $\sim$10000 \kms\ at +96d. What is striking, however, is the appearance
 of this feature in our second spectrum (+32d) where it has become very strong
 and broad. It is obvious that this absorption cannot be caused by Si only (it
 extends to negative velocities) but it must be a blend with more features.
 Attributing this to \Hei\ $\lambda$ 6678 is of course a possibility. However,
 this line would have to appear at very high velocities in order to appear
 blended with the Si line (in most SNe Ib, these two features are clearly
 distinguishable). In addition, the line would have to be very strong and that
 appears to be in conflict both with the velocity and strength of the only other
 \Hei\ line detected in this spectrum (\Hei\ $\lambda$ 7065) and with the sudden
 disappearance of this feature in the next spectrum. If we attribute the (now
 weak) feature on the right of \SiII\ $\lambda$ 6355 to \Hei\ $\lambda$ 6678 in
 the +58d spectrum, this would have to be at $\sim$5000 \kms\,, \ie\,, a velocity
 inconsistent with the one inferred by \Hei\ $\lambda$ 7065 at the same epoch.
 In general, we note that the appearance and disappearance of this deep trough
 at 6000 \AA\ is rather unusual in comparison to normal Type Ib events. 
\section {Distance and extinction toward \sn\,}
\label{sec:DistExt}
 To determine the bolometric light curve and physical properties of the
 transient, a correct estimation of the distance and extinction is essential.
 
 The host NGC 2770 is a well studied star forming galaxy with nine redshift
 independent distance estimations\footnote{http://www.ned.ipac.caltech.edu/}.
 Out of them, eight measurements used the `Tully Fisher' technique and for one
 IRAS  photometry was used. The measured values have a range between 26.1 Mpc
 and 36.0 Mpc, that corresponds to a weighted mean of 29.97$\pm$3.48 Mpc. The
 Hubble flow distance of the host\footnote{The
 cosmological model with $H_0$ = 70 \kms\, Mpc$^{-1}$,$\Omega_{m}$ = 0.3 and
 $\Omega_{\Lambda}$ = 0.7 is assumed throughout the paper and the uncertainty
 corresponds to a local cosmic thermal velocity of 208 \kms
 \citep{2002A&A...393...57T}.}, after correction for Virgo infall, is estimated
 as 29.3$\pm$2.1 Mpc. Combining the above measurements, we adopt the weighted
 mean distance of $29.5\pm1.8$ Mpc, which corresponds to a distance modulus of
 $\sim32.4$ mag.

 The Galactic reddening along the line of sight of \sn\ as derived from the
 100\mum\ all sky dust extinction map \citep{1998ApJ...500..525S} is \ebv\
 $=0.022\pm0.0004$ mag. However, measuring the total line-of-sight extinction
 towards \sn\ is non-trivial. \sn\ is associated with a large number (about 150)
 of \ha\ emitters \citep{2011AdSpR..47.1421G} which are expected to be star
 forming \Hii\ regions and presumably few of them can make the transient highly
 extinguished. The host is also a highly inclined spiral galaxy (see Table
 \ref{tab:propgal}), where the dusts are normally distributed along the edges of
 the young and old stellar discs (\citealt{2011A&A...527A.109P} and references
 therein) and hence expected to produce a large extinction toward the SN due to
 the foreground material.
 In fact, it has already been pointed out by \citet{2011ApJ...741...97D},
 with sufficient number of Type Ib events, that by and large the more extincted
 SNe reside in more inclined host galaxies. For \eg\,, SN 2007D is the most
 extinguished object which is hosted by a highly inclined ($\sim$70\degr\,)
 galaxy; however, there are exceptions like SN 2004ge.

 NGC 2770 hosted three SNe 1999eh, 2007uy and 2008D. The isophotal diameter of
 NGC 2770 is $d_{25} =$ 3.567\arcmin\,. This corresponds to an
 apparent radius $r_{25}$ = 1.734\arcmin\ of the host. On the other hand since
 NGC 2770 is a spiral galaxy, assuming the star forming regions and SNe sites
 are distributed on the plane of the disk, the measured values of the
 deprojected distances of SNe 2007uy, 2008D and 1999eh from the centre of the
 host are roughly 0.707\arcmin\, ($\equiv$5.3kpc), 1.210\arcmin\
 ($\equiv$9.1kpc) and 1.262\arcmin\ ($\equiv$9.5kpc) respectively. This implies
 $r_{2007uy} \sim 0.408r_{25}$, $r_{2008D} \sim 0.698r_{25}$ and $r_{1999eh}
 \sim 0.728r_{25}$. Thus all the SNe happened within half-light radius of the
 host. Moreover according to \citet{2009ApJ...698.1307T} the metallicities of
 these SNe sites are also comparable with each other (8.36$\pm$0.1, 8.53 and
 8.37 dex respectively for SN 2008D, \sn\,, SN 1999eh). So we can expect that
 the environments are similar and the host extinctions should have nearly equal
 values. 

 From the extinction values of the well studied nearby Type Ibc events,
 \citet{2011ApJ...741...97D} proposed an empirical relation to calculate the
 host reddening for Type Ibc events using the values of observed $V-R$ colour
 of the transient at +10d after $V$ and $R$ band maxima. According to this
 formalism, the magnitudes of the host reddening, \ebv\ for \sn\ measured w.r.t
 the $V$ and $R$ band maxima are respectively $0.79\pm0.18$ mag and
 $0.82\pm0.21$ mag.

 The colour excess of the host can be measured by means of Balmer
 decrement, usually from \ha\,/\hb\ ratio \citep{1989agna.book.....O} or from
 the
 equivalent width of the non-contaminated and non-saturated \Nai\,D absorption
 feature \citep{1990A&A...237...79B, 1994AJ....107.1022R, 2007snld.confE..11E,
 2012arXiv1206.6107P}.

 By measuring the Balmer decrement,
 \citet{2009ApJ...698.1307T} determined the \ebv\ $\sim1.4$ mag\footnote{This is
 important to mention that \citet{2009ApJ...698.1307T} did not calculate the
 \ebv\ exactly at SN position, rather at the locations which are roughly
 $\pm2\farcs5$ separations from SN position. This corresponds to more than
 350pc in physical size.} toward \sn\,. This corresponds to total visual
 extinction $\sim4.34$ mag
 considering the ratio of total-to-selective extinction (R$_V$) of 3.1. We note
 that with this value of visual extinction, the peak absolute magnitude of \sn\
 would be $\sim-21$ mag. This is roughly 3.4 mag brighter than the average peak
 magnitude of Type Ibc events \citep{2011ApJ...741...97D} and comparable with
 over-luminous events, though neither the photometric nor the spectroscopic
 evolution of this event matches with those of luminous SNe. It is needless to
 say that measurement of the \ha\,/\hb\ ratio is critically dependent on the
 correct flux calibration and prescriptions for deriving atmospheric and
 foreground extinction. The broad spectral lines of a bright SN is also a
 probable dominating source of error in determination of the ratios of underling
 lines. We perform the same exercise with +392d spectrum and calculate the
 reddening from Balmer decrement (see \eg\,\citealt{2013ApJ...763..145D}) using
 the expression \[\ebv = 1.97\times\,log_{10}[(\ha\,/\hb\,)_{obs}/2.86]~~.\] The
 derived value of reddening \ebv\,, is 0.62$\pm$0.06 mag.

 On the other hand, for \sn\ we found a prominent impression, {\tt F}, of the
 host \Nai\,D, though
 the two components are not resolved in the moderate resolution spectra. Using
 our four high S/N spectra (+32d, +58d, +96d and +122d) where the unresolved
 \Nai\,D feature was prominent, we calculate its mean equivalent width (EW) of
 $0.66\pm0.04$\AA\,. By using the expression of \citet{2012arXiv1206.6107P} for
 the unresolved \Nai\,D feature we derive the value of reddening to be
 \ebv\,$={0.08}^{+0.03}_{-0.02}$ mag. We note that this value is similar to
 the line of sight reddening towards the face-on or less inclined galaxies
 \citep{2009ApJ...696..713S, 2009ApJ...697..676S, 2011ApJ...736...76R} and too
 low for a transient hosted by a highly inclined galaxy having a substantial
 foreground intervening medium. This value is comparable to the lower limit
 of reddening (0.063) that can be inferred from \Nai\,D absorption feature in
 low resolution spectra \citep{2007snld.confE..11E}. However, in the
 same contribution it was also pointed out that there is no clear empirical
 relation between \ebv\ and \ew\,, at least in the regime of low-resolution
 spectroscopy. We also notice a prominent impression {\tt E} of
 \Hei\ $\lambda5876$ in the spectra of \sn\,, that is due to associated star
 forming region. This may be a prime source of contamination for the \Nai\,D.

 To avoid these problems, we measure the line ratio between \Nii\ $\lambda6583$
 and \ha\ and adopt an indirect method to measure the \ebv\ toward \sn\,.
 \Nii\,$\lambda6583$/\ha\ ratio is actually a good estimator for Oxygen
 abundance in extragalactic \Hii\ regions \citep{2004MNRAS.348L..59P}. After
 noticing a strong anti-correlation between \Nii\,$\lambda6583$/\ha\ ratio and
 \ha\ equivalent width, whereas strong correlations among \Oiia\,, \ha\,, \hb\,
 and \hg\ fluxes in a sample of about 74 star forming blue galaxies,
 \citet{2002A&A...396..503K} proposed an empirical relation between \ebv\ and
 \Nii\,$\lambda6583$/\ha\,. They found that since \Nii\, $\lambda6583$ and \ha\
 are two nearby lines, the strength ratio will be less affected due to galactic
 extinction while providing an approximate tracer of \ebv\ for highly reddened
 (\ebv\,$\ga0.1$ mag) extra-galactic objects. Note, though, this is inefficient
 for low reddened objects. The +392d spectrum, which has the least  SN
 contribution and mostly contains
 the emission lines due to \ha\ knots, has been used to calculate  \ebv\,.
 The measured value of \ebv\ by this relation is ${0.52}^{+0.20}_{-0.14}$ mag.
 This is similar to the reddening along highly reddened SNe 2004ge and 2007D
 and also close to the mean reddening (\ebv\,$\sim$ 0.65 mag) along SN 2008D
 \citep{2009ApJ...705.1139M}.
 
 For this work, we adopt the weighted average of four measurements,
 obtained from the empirical relations proposed by \citet{2002A&A...396..503K,
 2011ApJ...741...97D} and ratio of Balmer decrement. This implies that the host
 reddening toward \sn\ is $0.63\pm0.15$ mag. Since this is an order of magnitude
 higher than the galactic extinction, we anticipate that this is the total
 reddening along the line of sight of \sn\,. This corresponds to a total visual
 extinction (A$_V$) of $1.9\pm0.5$ mag.
 
\begin{figure*}
\centering
\includegraphics[width=17cm]{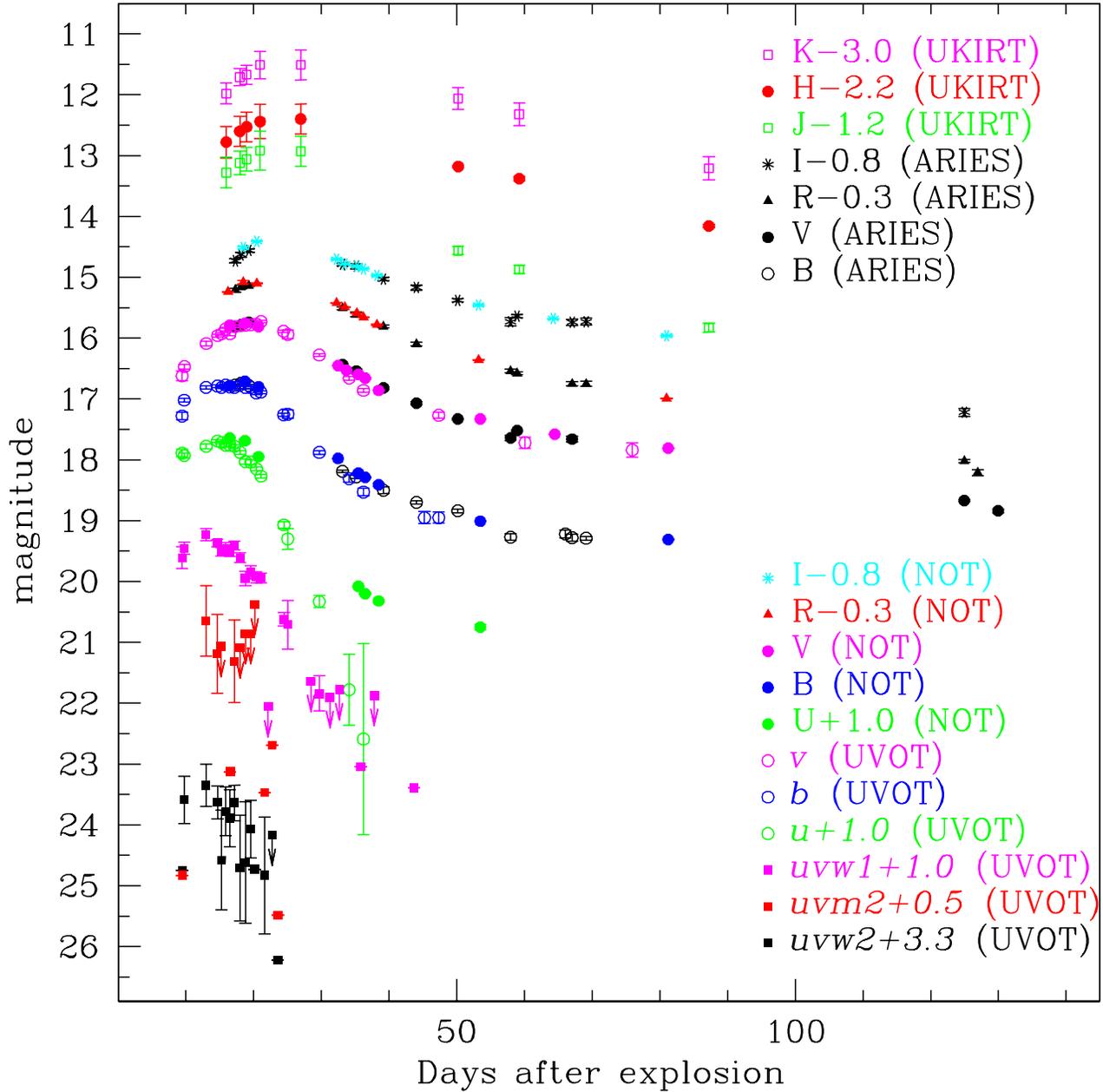}%
\caption{The light curves of SN 2007uy in NIR $JHK$, Optical $UBVRI$ and
 $uvw2, uvm2, uvw1 \& u, b, v$ NUV-Optical UVOT bands. The colour version of the
 figure is available in the online journal.}
\label{fig:appcur}
\end{figure*}

\section{Light curve evolution: NUV, Optical and NIR}
\label{sec:lightcurve}
 The calibrated light curves of \sn\ are presented in
 figure \ref{fig:appcur}. The data cadences of ground based Optical and NIR
 observations are $N(U,B,V,R,I,J,H,K)$ = (7,19,23,20,22,8,8,8), while that for
 spaced based NUV and Optical observations are $N(uvw2,uvm2,uvw1,u,b,v)$ =
 (14,8,18,19,28,30) respectively.
 Since the host redshift is small,  no `K-correction' is  applied. From the
 peak, the SN light is mainly powered by the radioactive decay of $^{56}$Ni and
 $^{56}$Co. The energy of $\gamma$-rays and positrons emitted by radioactive
 decay processes is fully absorbed by the SN ejecta and  re-emitted as 
 blackbody radiation. Hence in early phases Type I events can be approximated as
 `blackbody supernova' \citep{1982ApJ...253..785A}. The emergence of a young
 Type I SN with its blackbody nature is clearly noticeable in all bands. At a
 few phases, the SN was not detected in some of the
 $Swift$/UVOT bands for which the corresponding upper limits are mentioned.
 In the ground-based observations, the maximum sampling of SN light was done in
 the $V$ passband. Similarly, for the space-based $Swift$/UVOT observations the
 maximum sampling was done in $v$ passband. Since the central wavelength and
 bandwidth of the $V$ passband (5448\AA\ and 840\AA\ respectively) is fairly
 similar to that of the $v$ passband (5468\AA\ and 769\AA\ respectively), we
 have treated the combined data of the $V$ and $v$ bands as a single visual-data
 set for further analysis of photometric data and likewise the $B$ and $b$-band
 data sets have been combined as a single blue-band data set.
 In contrast, a significant deviation between the U and $u$ passband light
 curves, starting from early epochs ($\ga$20 days), is probably due to a
 considerable difference between their responses (for the U band: central
 wavelength 3663\AA\, bandwidth 650\AA\, for the $u$ band: central wavelength
 3465\AA\, bandwidth 785\AA\, respectively). 

 The evolution of  the SN light curve is composed of three distinct phases $-$
 (i)
 shock breakout phase (initial few days), (ii) photospheric phase (within first
 40$-$50 days) and (iii) nebular phase (beyond 50 days after explosion). For
 this particular event we will discuss the evolution of photospheric and nebular
 phase light curves.

\subsection{Evolution during photospheric phase}
\label{sec:firstepoch}
 The temporal variation of the visual light roughly resembles the temporal
 variation of the total UVOIR (NUV+Optical+NIR) light of the SN. The maxima of
 visual light also roughly corresponds to the epoch of maximum of the UVOIR
 bolometric light of the transient. The rise time of \sn\ in the visual band is
 19.4 day which is determined after estimating the maximum through third order
 polynomial fitting on the early part ($\la$40 days) of the visual light curve.
 This is comparable with the rise time derived for engine driven energetic, Type
 Ic SN 1998bw (16.14$\pm$0.08 days) and optically-normal Type Ib SN 2008D
 (18.82$\pm$0.24 days), whereas, it is longer than that of Type Ic SN 2002ap
 (10.12$\pm$0.20 days) and engine driven energetic SN 2006aj (10.37$\pm$0.14
 days).
 The measured rise time of \sn\ in $U$ band is 15.13$\pm$0.4 days, whereas in
 $I$ band it is 23.09$\pm$1.44 days. This shows that, similar to other Ibc
 events, \sn\ also evolved faster and peaked
 first at lower wavelengths and then at higher wavelengths.

 The peak width
 ($\Delta$d$_{0.25}$), defined as the width of the light curve, when the SN has
 faded by 0.25 mag from its maximum brightness, is also calculated for \sn\,.
 For the visual band it is roughly 11 days. This is comparable to the $V$ band
 peak width of SN 2006aj ($\Delta$d$_{0.25} \sim 11.1$ days) and SN 2002ap
 ($\Delta$d$_{0.25} \sim 10.6$ days), but less than engine driven SN 1998bw
 ($\Delta$d$_{0.25} \sim 13.3$ days) and normal Type Ib event SN 2008D
 ($\Delta$d$_{0.25} \sim 18.04$ days).

 The post maxima decay rates of \sn\ in the $V$ and $R$ bands are respectively
 0.058$\pm$0.002 mag.d$^{-1}$ and 0.042$\pm$0.002 mag.d$^{-1}$, which
 corresponds to post maxima decay parameters of ($\Delta$m$_{15}$) of
 0.87$\pm$0.003 and 0.63$\pm$0.03 mag respectively. Here $\Delta$m$_{15}$
 quantifies the decay in magnitude in 15 days after maxima. For the $B$ and $U$
 bands, the value of $\Delta$m$_{15}$ are even higher, around 1.2 and 2.1 mag
 respectively. This is in agreement with the prediction of models on SN
 evolution \citep{1999astro.ph..9034W}. The $V$ and $R$ band decay rates are
 comparable with the decay rates of Type Ibc events studied by
 \citet{2011ApJ...741...97D}. On the other hand, the spectroscopic study (see
 \S\ref{sec:specevo}) showed convincingly that the event neither belongs to the
 Type Ic nor to the class Ic-BL.
 
\begin{figure}
\centering
\includegraphics[width=8.5cm]{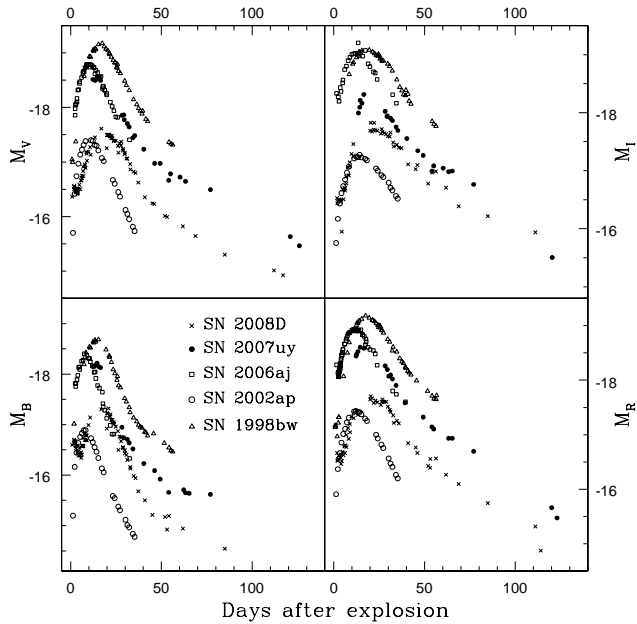}%
\caption{Absolute light curves of SN 2007uy. Comparison with other Type Ib and
 Ic events.}
\label{fig:abscur}
\end{figure}

 Adopting the distance modulus of \sn\ $\sim 32.4$ mag and total reddening along
 the
 line of sight \ebv\,$\sim$0.63 mag (see \S\ref{sec:DistExt}), we have computed
 absolute magnitudes in the $BVRI$ passbands to compare it with other Type Ibc
 events (figure \ref{fig:abscur}). For relatively redshifted  objects, like SN
 2006aj, the cosmological time dilation factor has been taken care into account
 to correct the corresponding time scale. In all bands \sn\ was intrinsically
 brighter than the normal Type Ib event SN 2008D and Type Ic event SN 2002ap.
 Though in the $R$ and $I$ passbands it was a little dimmer than the GRB/XRF
 associated events SNe 2006aj and 1998bw, but its peak absolute $R$ band
 magnitude
 $-18.5\pm0.16$ is about one mag brighter than the mean value ($-17.6\pm0.6$)
 derived for well observed Type Ibc events \citep{2011ApJ...741...97D}. In the
 visual and blue bands its peak magnitude is comparable with energetic events
 associated with GRB/XRFs. Hence, \sn\ is optically more luminous than normal
 Type Ib events and comparable to the engine driven energetic SNe.

\begin{figure}
\centering
\includegraphics[width=8.5cm]{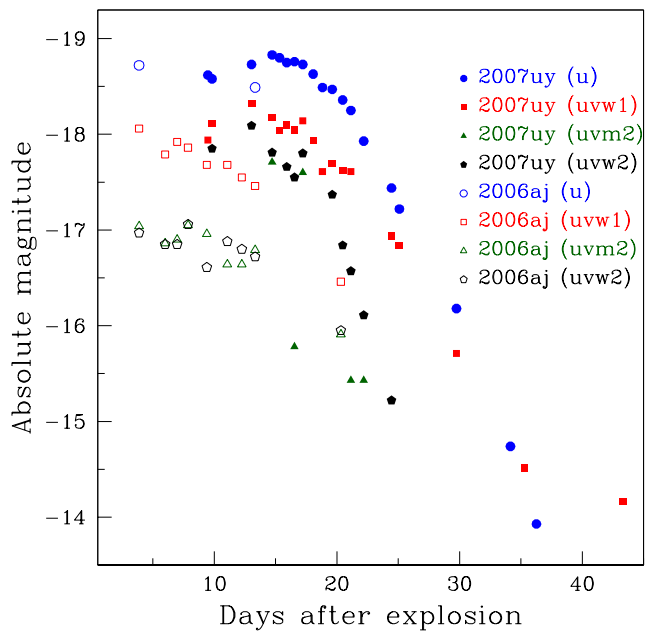}%
\caption{ NUV and u band absolute light curves of SN 2007uy. Comparison with
 Type Ic event SN 2006aj. Data for SN 2006aj were presented by
 \citet{2010A&A...523A..56S}. The colour version of the figure is available in
 the online journal.}
\label{fig:absuvot}
\end{figure}

 \sn\ was luminous in the NUV. It was detectable in the $Swift/uvw2$ and $uvm2$
 images until +24d, while in the $uvw1$ and $u$ filter systems it was detected
 until +44d. In contrast, SN 2008D which was more luminous in the X-ray and
 radio and happened in the same galaxy with almost similar environment
 \citep{2009ApJ...698.1307T}, was seldom detected in the $Swift$ NUV bands. As
 we have noticed that \sn\ is luminous in higher optical frequencies and it is
 comparable to engine driven energetic explosion like SN 2006aj, it is worth to
 compare the NUV flux of this event with that of engine driven SNe.
 Fortunately, $Swift$ has a good coverage of XRF060218/SN 2006aj in all
 NUV bands during the phases comparable to that of \sn\,. Figure
 \ref{fig:absuvot} shows a comparison

 between the NUV and $u$ absolute magnitudes of \sn\ with that of SN 2006aj. The
 UVOT data for SN 2006aj have been taken from \citet{2010A&A...523A..56S}. SN
 2006aj seems to be dimmer than \sn\,. The nature of the evolving supernova is
 clearly visible in all absolute light curves of \sn\,. In the case of SN
 2006aj, the nature of an evolving blackbody supernova is reflected in the $u$
 and marginally in the $uvw1$ light curves, whereas photometry of $uvm2$ and
 $uvw2$ images show a very shallow linear decay of SN light in the NUV range.
 The post maxima decay rates in these two bands, determined through a linear fit
 to these data sets, are respectively $0.06$ mag\,d$^{-1}$ and $0.04$
 mag\,d$^{-1}$. If all optical photons are
 generated through blackbody radiation under similar conditions of the ejecta,
 then post maxima decay rates in these two bands should be much higher than
 these values (assuming the SN will be optically thick during first 20 days);
 it was also predicted by \citet{1999ApJ...516..788W} through modeling of SN
 1998bw spectra in the spectral range between the $U$ and $I$ passbands. For SN
 2006aj, the measured post-maxima decay rate in the $I$ band is $\sim$
 0.05 mag\,d$^{-1}$ while in the $B$ and $uvw_1$ bands these are as high as
 $\sim$ 0.10 mag\,d$^{-1}$ and 0.12 mag\,d$^{-1}$ respectively. 
 Hence the decay rate in NUV bands should be higher. This discrepancy was
 also noted by \citet{2010A&A...523A..56S} and was explained as an effect of
 blackbody radiation emission \citep{2003LNP...598..113P} while the UV emitting
 area shrinks with time during the evolution of the transient.

\begin{figure}
\centering
\includegraphics[width=8.5cm]{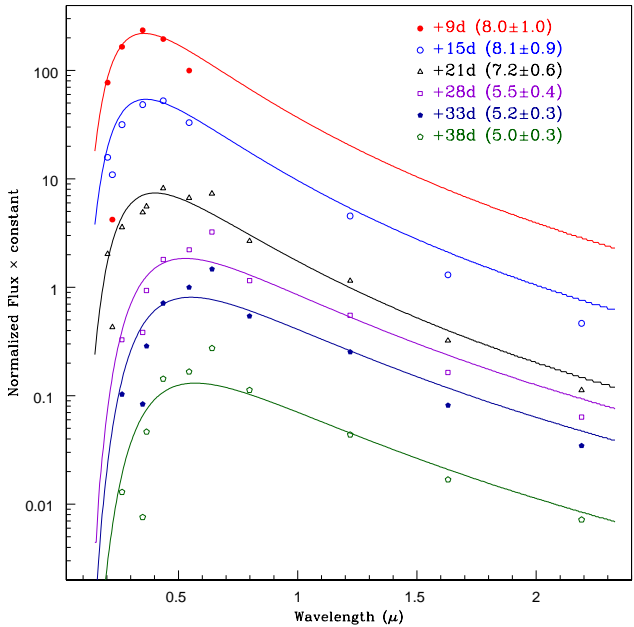}%
\caption{Temporal evolution of observed SED of \sn\,. The temperature of each
 blackbody is mentioned inside the parenthesis, with a temperature-scale 1000K.
 The one sigma uncertaities are measured by applying the `Monte Carlo' method.
 The colour version of the figure is available in the online journal.}
\label{fig:bb_7uy}
\end{figure}

 In contrast, blackbody fitting to the early  \sn\ data (figure
 \ref{fig:bb_7uy}) shows that envelope of this SN evolved more or
 less like a blackbody. Here we have plotted the flux, normalized with respect
 to the $V$ band flux, to nullify the effect due to time evolution of the
 angular
 diameter of the transient. The effect due to dilution of
 SN envelope is also neglected as a first approximation\footnote{It is worth
 noting that the measurement of the true Temperature of the SN envelope along
 with
 the measurement of its distance require a proper estimation of dilution factor
 through hydrodynamical modeling \citep{1974ApJ...193...27K,
 1996ApJ...466..911E}. This can only provide a colour temperature for the system
 along with a true estimation of distance independent of cosmology. The
 technique was properly applied by \citet{2001ApJ...558..615H,
 2002AJ....124.2490L} and others for Type IIP events. In the present work we
 provide a rough estimate of the overall temperature of the SN photosphere and
 adopted the value of distance obtained above (\S\ref{sec:DistExt}).}. The
 temperature of each blackbody is mentioned inside the parenthesis, with a
 temperature-scale 1000\,K. The one sigma uncertainties are
 measured by applying the `Monte Carlo' method. The initial
 temperature of the photosphere is $\sim 8000$\degree\,K. During the next 6 days
 the
 temperature slightly increased to $\sim 8100$\degree\,K and then started to
 decrease with time through a power law with temporal decay index $\sim-0.51$,
 which is comparable with the theoretical prediction ($-0.5$) for a `blackbody
 supernova' by \citet{1982ApJ...253..785A}.
 The initial increment in temperature, followed by a power law decay with time
 (decay index $\sim-0.12$) was also observed in the Type Ic SN 2002ap
 \citep{2004A&A...427..453V}. Thus we speculate that unlike XRF060218/SN 2006aj,
 \sn\ evolved roughly as a `blackbody supernova' from the very initial epoch.


\subsection{Evolution during nebular phase}
\label{sec:secondepoch}
 Normally after +50  to +60 days, the envelopes of most Type I events
 start to become optically thin, and hence the light curves start to flatten. In
 the $B$, $V$, $R$ and $I$ bands the decay rate of \sn\ is roughly $0.017 -
 0.018$ mag d$^{-1}$, whereas in $U$ band it is $\sim$ 0.03 mag d$^{-1}$. This
 decay rate is similar to the nebular decay rates of other Type Ibc events and
 slightly
 higher than the decay rate of the $^{56}$Co $\rightarrow$ $^{56}$Fe nuclear
 transition, which is typically 0.0098 mag d$^{-1}$. It is worth mentioning that
 in Type II SNe, mainly in normal Type IIP events, having a huge H shell in the
 outer surface, the nebular decay rates are comparable with that of $^{56}$Co
 (\citealt{2011ApJ...736...76R, 2011MNRAS.414..167R} and references therein).
 On the other hand, for low luminosity Type IIP events like SN 2005cs, the decay
 rate during nebular phase ($\sim 0.0046$ mag d$^{-1}$ in $V$ band during the
 period 140$-$320 days) is significantly lower than that of $^{56}$Co
 \citep{2009MNRAS.394.2266P}. This was also observed by
 \citet{2003MNRAS.338..939E} in the normal Type IIP event SN 1999em during its
 initial ($\sim$ one month) nebular phase. \citet{2007A&A...461..233U} explained
 this as an effect due to radiation generated inside the inner warmed ejecta,
 that propagates through the cool, optically thin outer region and makes the
 nebular light relatively stable. The relatively high decay in very late epoch
 (t\textgreater300 days) during the evolution of SN 2005cs was explained as 
 resulting either due to (i) dust formation in the SN ejecta, (ii) a lower
 efficiency of $\gamma$-ray trapping due to the decreased density of the ejecta,
 or due to (iii) cooling of inner ejecta \citep{2009MNRAS.394.2266P}.

 The optical and NIR nebular light curves of \sn\ show sharper decay
 ($\sim 0.02-0.03$ mag.d$^{-1}$) than the canonical decay of radioactive
 $^{56}$Co. A similar decay rate was also noticed for SN 2002ap
 \citep{2003MNRAS.340..375P, 2004A&A...427..453V}, which was explained as a
 leakage of $\gamma$-photons from the transparent and less massive SN
 atmosphere. Dust formation may also be another cause.
 With increasing time, the newly formed small dust grains in the ejecta can
 progressively reduce the fluxes of SN in all bands. Though there is a
 possibility for the creation of dust-grains of size $\la$ 0.01 \mum\ within the
 first 100$-$300 days in
 Type I SNe \citep{2008ApJ...680..568S, 2011ApJ...736...45N}, these hot grains
 (T$\sim$1600\degree\,K) would cause an IR excess in the SN continuum spectra.
 Moreover, dust formation should manifest itself by creating a blueshift in the
 line, increasing with time. Neither excess in the continuum, nor the temporal
 increment in the blueshifts of the spectral lines were observed for SN 2007uy
 (rather it decreases with time; see \S\ref{sec:specevo}). In fact beyond +50d
 \sn\ started to become bluer (see \S\ref{sec:ColBol}), which does not support
 dust formation during the period of observation. Thus the possibility of
 dust formation at the outer shell of the ejecta in early epoch is also unable
 to explain this deviation.

 Here we propose that leakage of $\gamma$-photons is the principal cause for
 rapid decay in light curve of \sn\,, like other stripped-envelope supernovae.
 We also admit that this simultaneously depends on the amount of ejected mass
 (M$_{ej}$) and  how fast it becomes diluted (optically thin). If E$_k$ is the
 kinetic energy of the explosion then the decay rate of the nebular phase is
 roughly proportional to M$_{ej}$/E$_k$$^{1/2}$ \citep{2008MNRAS.383.1485V}. So,
 explosions with higher ejecta mass and kinetic energy or lower mass with low
 kinetic energy should show a similar trend in nebular decay. For \eg\,, the
 very late time ($\textgreater 150$ days after explosion) evolution of SNe
 1998bw and 2006aj also show an almost identical decay rate in the $V,~R,~I$
 bands of $\sim 0.02$ mag.d$^{-1}$ and higher than the decay rate of $^{56}$Co
 \citep{2001ApJ...555..900P, 2011AIPC.1358..299M}. The first one is an explosion
 with higher mass of ejecta and kinetic energy, while the second shows a lower
 ejecta mass with smaller amount of kinetic energy.
 
\section {Evolution of colour and bolometric light}
\label{sec:ColBol}
\begin{figure}
\centering
\includegraphics[width=8.5cm]{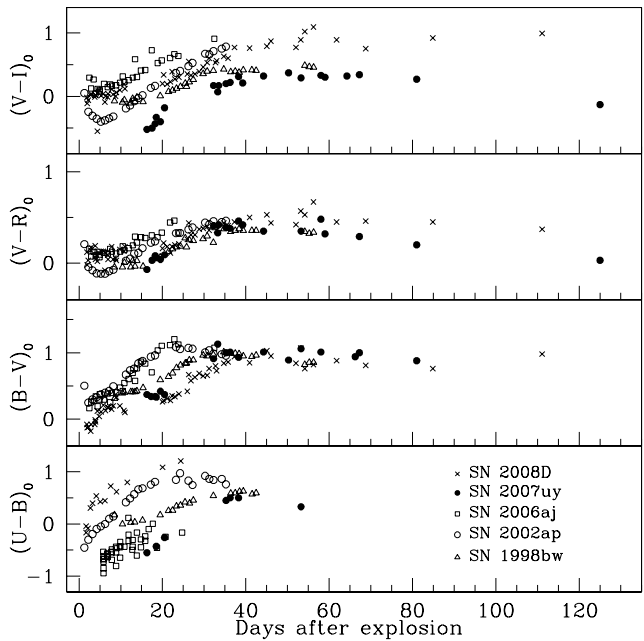}%
\caption{ Color evolution of SN 2007uy. Comparison with other Type Ibc events}
\label{fig:colour}
\end{figure}

 The reddening-corrected colour evolution of SN 2007uy is presented in
 figure
 \ref{fig:colour}. For comparison, the colour curves of the well-studied SNe
 1998bw \citep{1998Natur.395..670G}, 2002ap \citep{2003PASP..115.1220F}, 2006aj
 \citep{2006A&A...457..857F} and 2008D \citep{2008Natur.453..469S, roythesis}
 have been plotted. The evolution of intrinsic colour of \sn\ is similar to that
 of other events, however during early phases ($\la$+35d) the SN seems to be
 bluer than the other SNe, especially in $(V-I)_0$ colour. This is a
 manifestation of enhanced line emission due to \Hei\ $\lambda$5876 and \Oia\
 $\lambda$5577 in comparison to the higher wavelength part of the spectrum,
 which is also confirmed from +17d spectrum.
 Like other events, \sn\ also shows a transition from blue to red during its
 initial evolution ($\la$+40d) and turned bluish
 during its nebular evolution ($\ga$+50d). This is probably due to transition of
 ejecta from optically thick to optically thin state, which in turn initiates 
 the escape of high energy optical photons in a large amount from the SN
 ejecta.

\begin{figure}
\centering
\includegraphics[width=8.5cm]{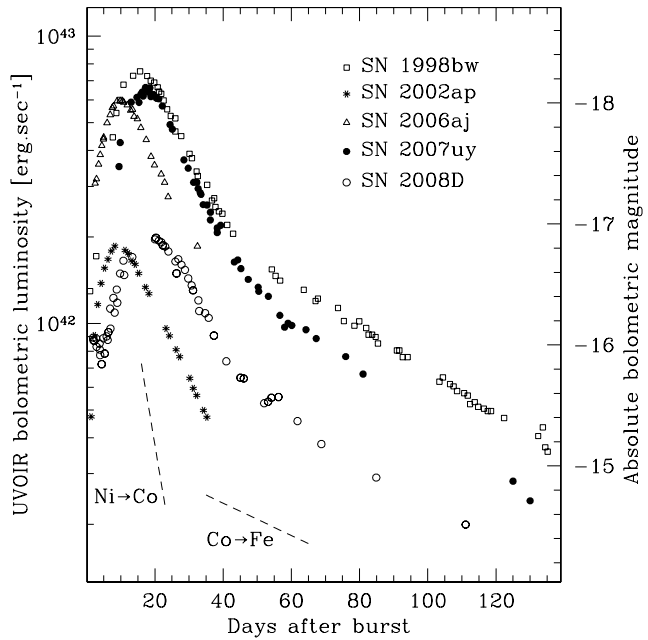}%
\caption{The UVOIR bolometric light curve of SN 2007uy. Comparison with other
 Type Ibc events. The colour version of the figure is available in the online
 journal.}
\label{fig:bol}
\end{figure}

 The bolometric light curve is crucial for the calculation of the synthesized
 radioactive $^{56}$Ni, the kinetic energy of the explosion, the amount of mass
 that was expelled during the explosion and hence to understand the nature of
 the progenitor.
 The quasi-bolometric UVOIR light curve has been computed from UV, optical and
 NIR data for the event that is presented in figure \ref{fig:bol}. Bolometric
 fluxes have been computed in those epochs for which 
 visual magnitudes are available. The extinction-corrected magnitudes are first
 converted into fluxes using zero points given by \citet{1998A&A...333..231B}
 and then the total flux in the UVOIR bands is obtained after a linear
 interpolation and integration between 0.203 and 2.19\micron. The $JHK$
 contribution during the photospheric phase is calculated from our data, whereas
 beyond +80d, in the nebular phase, due to lack of UV and NIR data, we are not
 able to make any direct measurement. For many Type Ia supernovae it has been
 observed that the NUV and NIR contribution at the nebular phase is roughly 20\%
 \citep{2000A&A...359..876C, 2009A&A...505..265L}. However, for Type Ic SN
 2002ap, \citet{2006ApJ...644..400T} found a relatively higher contribution
 ($\sim25-30$\%) from NIR light during comparable epochs. So, for present
 analysis, beyond +80d, we increase the bolometric flux by 35\% to estimate the
 maximum possible contribution from NUV and NIR light and construct the UVOIR
 quasi-bolometric light curve. Similarly, before the peak, there are few epochs
 for which we do not have any $RI$ and $JHK$ data. To calculate the magnitudes
 at those epochs, a blackbody spectrum has been fitted to construct the SED for
 each day (see figure \ref{fig:bb_7uy}) and synthetic magnitudes have been
 calculated from the SED. 

 The peak UVOIR bolometric luminosity of \sn\ is about 6.4$\times10^{42}$
 erg s$^{-1}$ which is comparable to that of the engine-driven energetic SN
 1998bw ($\sim 7.5\times10^{42} {~\rm erg s}^{-1}$) and SN 2006aj ($\sim
 5.9\times10^{42} {~\rm erg s}^{-1}$), and much higher than the broad-lined Type
 Ic event SN 2002ap ($\sim 2.1\times10^{42} {~\rm erg s}^{-1}$) and the
 optically-normal Type Ib event SN 2008D ($\sim 2.0\times10^{42}
 {~\rm erg s}^{-1}$). The peak luminosity of Type I SNe is roughly proportional
 to the ejected amount of $^{56}$Ni because the contributions from shock heating
 and CSM interaction are small. Since the maximum light of \sn\ is almost
 similar to that of SNe 2006aj and 1998bw, hence the amount of synthesized
 radioactive $^{56}$Ni nuclei should be comparable with these engine driven
 events. On the other hand, the decay rate in nebular phase is comparable with
 decay of $^{56}$Co to $^{56}$Fe, though there is a slight deviation as
 mentioned in \S\ref{sec:secondepoch}.

 The bolometric light of a CCSN can comprise light from different components,
 such as the explosion of a single massive progenitor, the emergence
 of a magnetar, the interaction of ejecta with the circumstellar medium (CSM),
 or possibly due to a combination of these. The signature of SN
 shock-interaction with its CSM can be seen through radio observations (see
 \S\ref{sec:modeling}).

\section{Physical parameters of the explosion}
\label{sec:modeling}
 To determine the physical parameters of the exploding star, modeling of the
 UVOIR bolometric light is essential. In addition to that, to quantify the
 interaction with circumstellar medium, modeling of the radio data is also
 required.
\subsection{Modeling of optical light curve}
\label{sec:modeling_opt}
 To derive the physical parameters, we have followed the methodology of
 \citet{2008MNRAS.383.1485V}, \ie\ the early and nebular phase have been
 modeled separately.

\begin{figure}
\centering
\includegraphics[width=8.cm, angle=0]{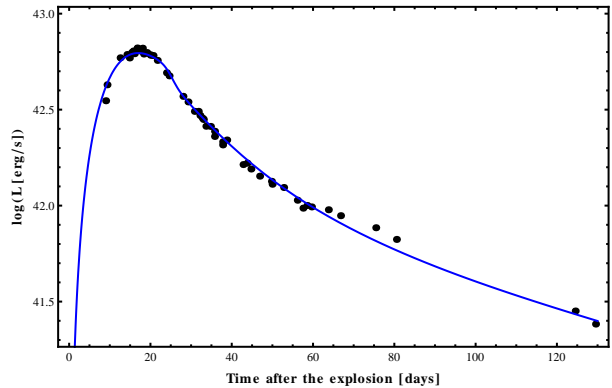}%
\caption{The modeled UVOIR bolometric light curve of SN 2007uy. The black dots
 are the observed data points. The blue solid line represents the model. The
 colour version is available in online journal.}
\label{fig:opt_model}
\end{figure}

 For $t \la 30$ days past explosion, the ejecta is presumed
 to be optically thick, having a spherically symmetric,
 homologous expansion and all the
 radioactive $^{56}$Ni is located in the centre. The model also
 assumes that the radius of the expanding shell is much higher than the
 radius of the progenitor, the optical opacity ${\kappa}_{opt}$
 $\sim 0.06$ cm$^2$g$^{-1}$
 remains constant through out its evolution \citep{2003ApJ...593..931M}, and the
 total radioactive energy of $^{56}$Ni $\rightarrow$ $^{56}$Co $\rightarrow$
 $^{56}$Fe is responsible for the heating of the ejecta and the diffusion
 approximation is valid for the photons \citep{1982ApJ...253..785A,
 2008MNRAS.383.1485V, 2011MNRAS.411.2726B, 2011ApJ...728...14P,
 2012A&A...539A..76O}.
 In the nebular phase (beyond $\ga 60$ days), the ejecta
 become optically thin and the emitted luminosity is powered by three sources
 $-$ instantaneous energy of the $\gamma$-rays generated during the $^{56}$Co
 decay, energy of the $\gamma$-rays coming from electron-positron
 annihilation and  due to the kinetic energy of the positrons
 \citep{1984ApJ...280..282S, 1997A&A...328..203C}.
 The low and high $\gamma$-ray trapping, respectively at early and late phases,
 have been invoked by dividing the ejecta into two components $-$ a high-density
 inner region and a low-density outer region \citep{2003ApJ...593..931M}.
 The outer region dominates the
 total emission mechanism at early epochs, \ie\ in the optically thick regime,
 and emission from the inner region, with higher $\gamma$-ray opacity, dominates
 the nebular phase.

 Optical modeling, as mentioned above, returns the values of four free
 parameters used to fit the bolometric light curve of \sn\,. These are: the
 total mass of the ejected material (M$_{ej}$), the total mass of $^{56}$Ni
 produced in the envelope (M$_{Ni}$), the mass fraction of the inner ejecta
 component ($f_{\rm M}$) and the fraction of kinetic energy contributed by the
 inner component ($f_{\rm E}$).
 To break the degeneracy between kinetic energy and ejected mass, the expression for velocity at peak luminosity has been used (\citealt{1982ApJ...253..785A};
 also see the equation 2 of \citealt{2008MNRAS.383.1485V} and equation 3 of
 \citealt{2012A&A...539A..76O}).
 
 Velocities of the ejecta are measured from the blue shifts of the absorption
 dips of the P-Cygni profiles, but due to the heavy line blanketing
 identification of
 photospheric lines is not trivial and also the possibility to get the
 undisturbed lines is much less. Our first spectrum, taken near maximum,
 contains the \Hei\ $\lambda$5876 line, with a considerable blue-shifted
 absorption
 dip. The measured velocity corresponding to this blue shift is roughly 15200
 km~s$^{-1}$. Assuming the He shells to be marginally detached from the
 photosphere, we can constrain the upper limit of the photospheric velocity
 ($v_{ph}$) within the above value. 

 The fitted model is presented in figure \ref{fig:opt_model}.
 The parameters obtained after modeling the observed bolometric light of this
 event are as follows : M$_{ej} \sim$ 4.4\msun\,, M$_{Ni} \sim$
 0.3\msun\,, $f_{\rm M} \sim$ 9\% and $f_{\rm E} \sim$ 0.02\%. Since the peak
 velocity and total ejected mass are related to the Kinetic energy
 (E$_{\rm K}$) of the entire ejecta by the expression:
\[ {\rm E}_{\rm K} = [(5/6)~{\rm M}_{ej}~(1-f_{\rm M})~v_{ph}^2]/(1-f_{\rm E}),\] the kinetic energy of the ejecta turned out to be E$_{\rm K} \sim
 15\times10^{51}$ ergs.

\subsection{Modeling of radio data}
\label{sec:modeling_radio}
 The radio light curve can be explained in terms of the evolution of the ejecta,
 which further determines the shape of the spectrum and evolution of its peak
 with time \citep{1998ApJ...497L..17S, 1999ApJ...523..177W}. It can also be
 explained by
 Synchrotron Self Absorption (SSA) with high brightness temperature and/or by
 a Free Free Absorption (FFA) from the ambient medium, which produces an
 exponential cutoff below the peak. A decrease in absorption with time,
 resulting a smooth, rapid turn-on, first at higher frequencies and later at
 lower frequencies. For each frequency, the peak flux corresponds to optical
 depth $\sim$ 1 and a power-law decline in the light curve can be seen.
 Finally, the spectral index approaches a constant negative value, which
 corresponds to non-thermal radiation in an optically thin medium
 \citep{1986ApJ...301..790W, 1990ApJ...364..611W, 1998ApJ...499..810C} and
 \citet{1986ApJ...301..790W} proposed a methodology for parameterization of
 SN radio light curve. 

\begin{figure}
\centering
\includegraphics[width=8.5cm]{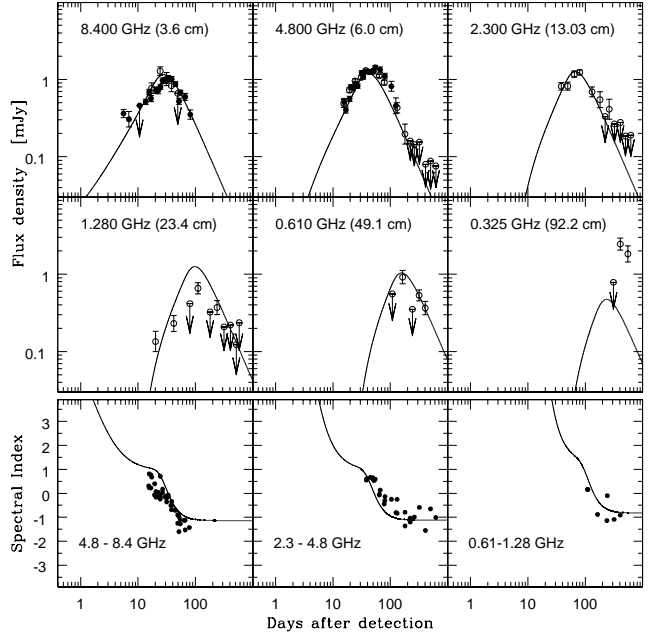}%
\caption{The modeled radio light curve of SN 2007uy. The upper panel shows the
 variation of radio signal with time. Open circles are the data obtained from
 \citet{2011ApJ...726...99V}, whereas the filled circles are VLA data
 acquired for the present work. The solid curves represent the best fit model.
 The lower panel shows the variation of spectral index with time.}
\label{fig:radio_model}
\end{figure}

 For the present work, we have parameterized the radio light curves of \sn\ to
 calculate the mass loss rate. The Upper panel of figure \ref{fig:radio_model}
 presents the results of radio observations in six different frequencies. The
 early time data is too sparse to determine the spectral index ($\alpha$) for
 the early part of the light curve, during the optically thick regime. Radio
 follow-up of this event has already been reported by
 \citet{2011ApJ...726...99V}.

  \begin{table}
  \caption{The parameters derived from the fitting of radio data}
  \label{tab:radio_model} 
  \centering
  \begin{tabular}{lc} \hline \hline
     \noalign{\smallskip}
      Parameters& Value\\ 
     \noalign{\smallskip} \hline
     \noalign{\smallskip}
     K$_1$&1500.75\\
     $\alpha$&$-1.1435$\\
     $\beta$&$-1.734$\\
     K$_2$&91.38\\
     t$_0$&$-3.59$ days\\
     $\delta$&2.41\\
     K$_3$&5.408$\times10^5$\\
     K$_4$&0.0036\\
     ${\delta}'$&$-3.43$\\
     K$_5$&1630.03\\
     ${\delta}''$&$-4.60$\\
     K$_6$& $\sim 1.0008$\\
     ${\delta}'''$&$-4.20$\\
     \noalign{\smallskip}
     \hline
  \end{tabular}
  \newline 
  \end{table}

 To model the light curves we have also included the literature data along with
 this new data set. The modeled parameters are reported in Table
 \ref{tab:radio_model}. The
 notation used for the parameters is as defined by \citet{2011ApJ...740...79W}.
 The details for the modeling can be found in \citet{1986ApJ...301..790W,
 1990ApJ...364..611W, 2011ApJ...740...79W}. 

 Out of the six frequencies of the data set, only three  (8.4, 4.8
 and 2.3 GHz) have been used to model the radio data through simultaneous
 fitting, while for other frequencies (1.28, 0.610 and 0.325 GHz) the expected
 theoretical light curves have been plotted along with observational data. Our
 best fitting model gives the value of overall reduced-${\chi}^2 = 7.14$. This
 is relatively higher than the value (4.42) mentioned by
 \citet{2011ApJ...740...79W} for the well-studied Type Ibc radio SN 1994I. This
 higher value is presumably a reflection of a lack of data due to less coverage
 of \sn\ in radio band. The lower panel of figure \ref{fig:radio_model} presents
 the temporal evolution of $\alpha$. The epochs for which we have an
 observation in a given frequency, the probable flux for other frequencies was
 determined by interpolation through polynomial fitting of the observed
 light curve. The thick curve shows that initially the source was extremely
 optically-thick with high +ve value of spectral index which turned to a
 constant -ve value $\sim -1$ asymptotically, that clearly marks the transition
 from a optically thick to thin state. A lower value ($\sim 2.0$) of $\alpha$
 than 2.5 in the optically thick regime shows FFA is also responsible for the
 initial radio flux along with the SSA. Although, the coefficient K$_6 =
 1.00085$, which is associated with FFA is also sufficiently less than the
 coefficient K$_5 = 1630.031$ responsible for SSA, the temporal evolution of
 both processes are almost similar (${\delta}'' = -4.602$ and ${\delta}''' =
 -4.206$ respectively). Thus the initial emission is mostly due to SSA in
 optically thick ejecta, though a mild contribution due to FFA can not be ruled
 out. This result is consistent with the conclusion drawn by
 \citet{2011ApJ...726...99V}.

 The modeling of the three frequencies shows that the radio flux of the source
 is moderate (K$_1 = 1500.75$) with a temporal decay index ($\beta$) roughly
 $= -1.73$ while the overall spectral index ($\alpha$) is $\sim -1.14$,
 comparable with the optically thin condition. It predicts that the object was
 discovered about (t$_0$) $= -3.6$ days after the explosion, which is consistent
 with the optical follow-ups. Following \citet{1982ApJ...259L..85C,
 1982ApJ...259..302C}, we assume that the temporal decrement ($\delta$) of the
 opacity of a uniform circumstellar medium (CSM)
 is roughly dependent on $\alpha$ and $\beta$ by the relation $\delta = \alpha -
 \beta -3$. A sufficiently lower value of K$_2$ (= 91.38), than K$_3$ (=
 $5.408\times10^5$) implies that the absorption processes in CSM are mainly due
 to its clumpy/structured form than its uniformity. On the other hand a
 relatively higher value of ${\delta}'$ (= $-3.43$) than $\delta$ (= $-2.41$)
 implies that the effect of clumpiness will not sustain for a longer time in
 comparison to the absorption due to uniform CSM. So, eventually the CSM starts
 to behave like a circumburst medium having a statistically small number of
 clumps. The lower value of the term K$_4$ (= $3.6\times10^{-3}$) demands that
 the radio absorption due to FFA in distant \Hii\ regions, which are in between
 the SN and us is quite low. In fact this is one order less than other events
 likes SNe 1978K, 1994I and 1998bw, but higher than SN 1993J
 \citep{2002ARA&A..40..387W, 2011ApJ...740...79W}. This is in contrast with the
 large extinction toward \sn\,, derived from optical data (\S\ref{sec:DistExt}).

 Finally, we find the mass loss rate during the pre-SN phase using radio data,
 adopting the prescription of \citet{2001ApJ...562..670W}. Assuming 50\% of the
 CSM along the line of sight is clumpy and the size of a clump is roughly 1/3 of
 the dimension of the expanding SN shock, then the volume filling factor by the
 blast wave is $ \phi \sim 0.22$. The effective optical depth will be given by
 $\left<{\tau}^{0.5}_{eff}\right> \sim 0.204$ (using equations 5 and 16 of
 \citealt{2001ApJ...562..670W}). Assuming the wind velocity of the progenitor
 $\sim 10$ km~s$^{-1}$ (the least possible value if the progenitor is a red
 supergiant) and initial velocity ($v_s$) and electron temperature ($T_s$) of
 the
 shock are respectively of the order of $20,000$ km\,s$^{-1}$ and 20,000 K, we
 can quantify the mass loss rate by:
\[\dot{M} = 6.79\times10^{-7}~(w_{wind}/10 {\rm km~s}^{-1}).(v_s/20,000 {\rm km~s}^{-1})^{1.5}.\]
 $~~~~~~\times(T_s/20,000{\rm K})^{0.68}$ \msun\,~yr$^{-1}$\\\\

 Since the photospheric velocity and temperature, measured from spectroscopy and
 photometry at around optical peak are roughly 15,200 km~s$^{-1}$ and 8,000
 K\footnote{This is important to mention that this is the colour temperature of
 the photosphere and hence does not correspond to the electron temperature of
 the source but provides an approximate estimation of the source temperature.}
 respectively, then the mass loss rate should be $\dot{M} \ga
 2.41\times10^{-7}~(w_{wind}/10 {\rm km~s}^{-1})$ \msun\,~yr$^{-1}$. This value
 is comparable with the pre-SN mass loss rate for other Type Ibc events derived
 from radio data \citep{2011ApJ...740...79W}. Observations and theoretical
 studies predict that the progenitors of Type Ibc are WR stars, having extremely
 high wind velocity $\ga 10^3$ km s$^{-1}$ \citep{2012A&A...544L..11Y,
 1996Ap&SS.237..145W}. Then the mass loss rate of the progenitor of \sn\ during
 its WR phase was roughly $\ga 2.41\times10^{-5}$ \msun\,.yr$^{-1}$. Certainly
 this result is consistent with other Type Ibc events
 \citep{2011ApJ...740...79W}.

\section{Conclusions}
\label{sec:conclu}
 Since the detection of the progenitors of Type Ibc events in pre-SN images is
 a technically challenging task, rigorous observational follow-up along with
 theoretical interpretation is necessary to understand these catastrophic
 explosions in a better way.

 In this article, we have presented an extensive follow-up campaign of a normal
 Type Ib event \sn\,,
 using the data acquired in a wide range of wavelength $-$ from NUV to radio
 bands along with optical spectroscopy. The radio modeling predicts that the
 event was discovered within 4 days after the explosion, while the early
 spectral features constrain the time difference between explosion and detection
 within 7 days. We have adopted the former value and assumed
 that the SN was discovered within 4 days after the explosion. The main results
 are summarized below.

 The extinction (A$_V = 1.9\pm0.5$ mag) along the line of sight of \sn\ is
 mainly dominated by the host galaxy. The highly inclined host and the presence
 of numerous \ha\ emitters along the line of sight make the transient highly
 extincted.

 The evolution of \sn\ was like a canonical `blackbody supernova'. The UV
 component is probably not due to shock breakout, rather it is a manifestation
 of radioactively heated, relatively less massive ejecta. As the ejecta become
 progressively optically thin, the optical photons start to escape rapidly,
 causing a steepening in the late time optical light curve. 

 The spectral evolution of \sn\ is faster than other Type Ibc events. Most of
 the lines get diluted quickly in comparison to other archetypal events. Early
 evolution of most of the lines is highly aspheric, though their natures are
 different from each other. This demonstrates that there is no dependency of
 asphericity on the distribution of radioactive $^{56}$Ni inside the ejecta, at
 least for this particular case. We propose that \sn\ is an aspheric explosion,
 which in due course of time attained a large-scale spherical symmetry.

 The nature of colour evolution of \sn\ is similar to other stripped-envelope
 SNe though this transient is intrinsically more bluish than other Type Ibc
 events. \sn\ is one of the luminous Type Ib events in respect to its UVOIR
 bolometric flux. The peak flux is comparable with the engine driven SNe 1998bw
 and 2006aj.

 The basic parameters of the explosion have been derived from the optical and
 radio
 data modeling. The UVOIR optical data has been modeled assuming the event as a
 `point explosion' after adopting the formalism of \citet{1982ApJ...253..785A},
 along with the modifications introduced by \citet{2008MNRAS.383.1485V}. Optical
 modeling predicts that about 4.4\msun\ was ejected with an explosion energy
 $\sim 15\times10^{51}$ erg and roughly 0.3\msun\ was produced during this
 explosion. The radio data shows a SSA dominated light curve evolution of \sn\,,
 though the contribution of FFA during the early phase can not be ruled out. The
 line of sight is probably dominated by relatively cool gas and dust making the
 optical extinction high, but due to lack of free-electron content, absorption
 of radio waves through FFA is less. It is one order less than that for other
 radio SNe. The pre-SN mass loss rate obtained from the radio data is
 $\dot{M} \ga 2.03\times10^{-7}~(w_{wind}/10 {\rm km~s}^{-1})$
 \msun\,~yr$^{-1}$. This value is consistent with the results obtained for other
 Type Ibc radio SNe, though an order of magnitude lower than typical value of
 mass-loss rate derived from X-ray. Assuming a WR star as a progenitor with
 wind velocity $\ga 10^3 {\rm km~s}^{-1}$, we can constrain the lower limit of
 pre-SN mass loss rate at $\dot{M} \ga 2.03\times10^{-5}$ \msun\,~yr$^{-1}$.

\section*{Acknowledgments}
 We thank all the observers at ARIES who provided their valuable time and
 support for the observations of this event. We are thankful to the observing
 staffs of NOT, UKIRT, NTT, MMT and VLT for their kind cooperation in
 observation of \sn\,. This work is based on the data obtained from the ESO
 Science Archive Facility. For this research work the VLA radio data has been
 used. The VLA is operated by the National Radio Astronomy Observatory, a
 facility of the National Science Foundation operated under cooperative
 agreement by Associated Universities, Inc. DARK is funded by the DNRF.
 Giorgos Leloudas is supported by the Swedish Research Council through grant No.
 623-2011-7117. Rupak Roy is thankful to the COSPAR fellowship program under
 which, he got the opportunity to visit DARK and initiated the entire project.
 S. B. Pandey and Rupak Roy acknowledge DST-RFBR grants INT/RFBR/P-25
 (2008-2010) and INT/RFBR/P-100 (2011-2013) for the present work.
 This research has made use of $Swift$/UVOT data obtained through the High
 Energy Astrophysics Science Archive Research Center (HEASARC) Online Service,
 provided by the NASA/Goddard Space Flight Center. The authors would also like
 to thank the anonymous referee for the comments and suggestions which helped in
 improvement of this manuscript.

\label{lastpage}
\end{document}